\documentclass[%
 reprint,
  showpacs,
 amsmath,amssymb,
 aps,
 prb,
 superscriptaddress,
 nobibnotes
]{revtex4-1}

\usepackage{graphicx}% Include figure files
\usepackage{dcolumn}% Align table columns on decimal point
\usepackage{bm}% bold math
\makeatletter
\AtBeginDocument{\@ifpackageloaded{natbib}{\ifNAT@numbers\if@filesw\immediate\write\@auxout{\string\global\string\NAT@numberstrue}\fi\fi}{}}
\makeatother
\usepackage{amsmath}

\begin{document}

%\preprint{APS/123-QED}

\title{Quasiparticle Interference in Fe-based Superconductors Based on a Five-Orbital Tight-Binding Model}
% Force line breaks with \\

\author{Youichi Yamakawa}
 \email[]{yamakawa@s.phys.nagoya-u.ac.jp}
\affiliation{Department of Physics, Nagoya University, Furo-cho, Nagoya 464-8602, Japan }
% \author{Seiichiro Onari}
% \affiliation{Department of Applied Physics, Nagoya University and JST, TRIP, Furo-cho, Nagoya 464-8603, Japan}
\author{Hiroshi Kontani}
\affiliation{Department of Physics, Nagoya University, Furo-cho, Nagoya 464-8602, Japan }

% %\collaboration{MUSO Collaboration}%\noaffiliation

% \author{}
% %  \homepage{http://www.Second.institution.edu/~Charlie.Author}
% \affiliation{
%  .
% }%
% % \affiliation{
% %  Third institution, the second for Charlie Author
% % }%

\date{\today}% It is always \today, today,

\begin{abstract}
We investigate the quasiparticle interference (QPI) in Fe-based superconductors
 in both the $s_{++}$-wave and $s_{\pm}$-wave superconducting states
 on the basis of the five-orbital model. 
In the octet model for cuprate superconductors with $d_{x^2 - y^2}$-wave state, 
 the QPI signal due to the impurity scattering
 at $\bm{q} = \bm{k}_i - \bm{k}_j$ ($E = | \Delta ( \bm{k}_i ) |$, $i = 1 \sim 8$) disappears
 when the gap functions at $\bm{k}_i$ and $\bm{k}_j$ have the same sign. 
However, we show that this extinction rule does not hold in Fe-based superconductors with fully-gapped $s$-wave state. 
The reason is that the resonance condition $E = | \Delta ( \bm{k}_i ) |$ is not satisfied
 under the experimental condition for Fe-based superconductors. 
We perform the detailed numerical study of the QPI signal using the $T$-matrix approximation, 
 and show that the experimentally observed QPI peak around $\bm{q}_2 = ( \pi, 0 )$ can be explained on the basis of
both the $s_{++}$-wave and $s_{\pm}$-wave states. 
Furthermore, we discuss the magnetic field dependence of the QPI by considering the Zeeman effect,
 and find that the field-induced suppression of the peak intensity around $\bm{q}_2$ can also be explained
 in terms of both the $s_{++}$-wave and $s_{\pm}$-wave states. 

%  \begin{description}
% Secondary publications and information retrieval purposes.
% \item[PACS numbers]
% May be entered using the \verb+\pacs{#1}+ command.
% You may use the \texttt{description} environment to structure your abstract;
% use the optional argument of the \verb+\item+ command to give the category of each item.
%  \end{description}
\end{abstract}

% 74.70.Xa 	Pnictides and chalcogenides 
% 77.80.B- 	Phase transitions and Curie point (for Curie point in ferromagnetic materials, see 75.30.Kz)
% 61.50.Ks 	Crystallographic aspects of phase transformations; pressure effects (see also 81.30.Hd in materials science)
% 74.25.Jb 	Electronic structure (photoemission, etc.) 
% 71.18.+y 	Fermi surface: calculations and measurements; effective mass, g factor
% 75.25.Dk 	Orbital, charge, and other orders, including coupling of these orders 
% 74.25.Kc 	Phonons
% 74.55.+v 	Tunneling phenomena: single particle tunneling and STM
% 74.20.-z 	Theories and models of superconducting state
% \keywords{Suggested keywords}%Use showkeys class option if keyword
\maketitle

%\tableofcontents

%%%%%%%%%%%%%%%%%%%%%%%%%%%%%%%%%%%%%%%%%%%%%%%%%%%%%%%%%%%%%%%%%%%%%%%%%%%%%%%%%%%%%%%%%%%%%%%%%%%%%%%%%%%%%%%%%%%%%%%%
\section{Introduction}
\begin{figure}[htb]
	\includegraphics[scale=0.4]{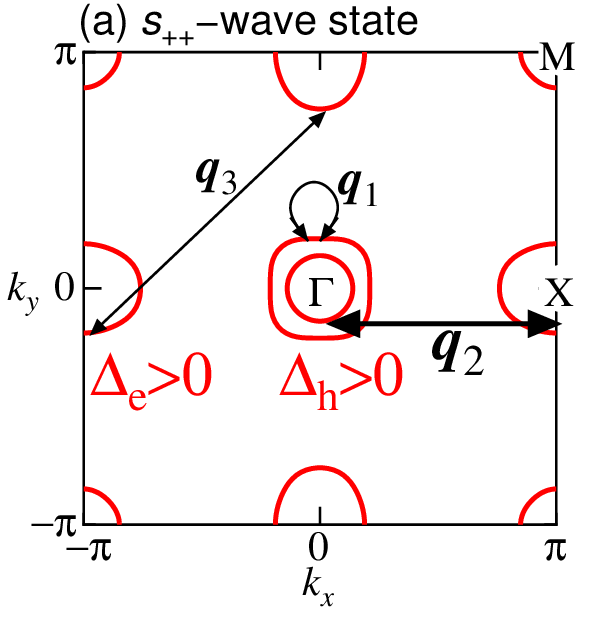}
	\includegraphics[scale=0.4]{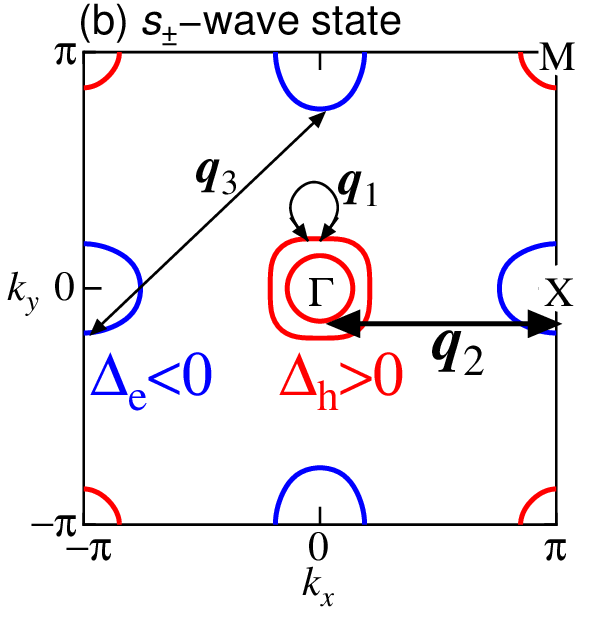}
	\caption{(Color online)
		Fermi surfaces and gap structures in the (a) $s_{++}$-wave and (b) $s_{\pm}$-wave states. 
		The arrows denote scattering wave vectors. 
		The scattering vector $\bm{q}_2 \sim ( \pi, 0 )$,
		 which is equal to the nesting vector,
		 connects hole and electron Fermi pockets. 
		$\bm{q}_1 \sim ( 0, 0 )$ corresponds to the intra-band scattering,
		 and $\bm{q}_3 \sim ( \pi, \pi )$ corresponds to the scattering within electron or hole Fermi pockets. 
	}
	\label{fig1}
\end{figure}

Since the discovery of Fe-based superconductors, \cite{Kamihara}
 much effort has been devoted to reveal the mechanism of high-$T_{\rm c}$ superconductivity (SC). 
The mother compounds exhibit structure and antiferromagnetic transitions.
These transitions are suppressed by carrier doping and then the SC state emerges. 
In the early theoretical studies, spin fluctuation mediated $s_{\pm}$-wave state,
 in which the SC gap functions change their sign between the hole and electron Fermi surfaces (FS),
 was proposed. \cite{Kuroki, Mazin, Chubukov, Graser, Hirschfeld}
On the other hand, the orbital fluctuations can induce the $s_{++}$-wave state
 without sign change in the gap functions as discussed in Refs.~\onlinecite{Kontani, Saito, Onari}. 
Figure~\ref{fig1} shows the unfolded FS and schematic picture
 of the (a) $s_{++}$-wave and (b) $s_{\pm}$-wave states. 
The $s_{++}$($s_{\pm}$)-wave state is driven by the orbital (spin) fluctuations at $\bm{q}_2 = ( \pi, 0 )$
 that corresponds to the nesting between hole and electron FSs. 

To distinguish between the $s_{\pm}$-wave and $s_{++}$-wave states,
 various phase sensitive experiments have been performed, 
 such as
 the impurity effect on $T_{\rm c}$, \cite{Sato, Li, Nakajima}
 the resonant peak by the inelastic neutron scattering, \cite{Inosov}
 the coherence peak by the nuclear magnetic resonance, \cite{Sato, Nakai}
 the quasiparticle interference (QPI)
 by the scanning tunneling microscope (STM), \cite{Hanaguri, Hanaguri_arxiv, Chi} and so on. 
Many theorists have preformed theoretical investigations of such experiments
 based on the realistic five-orbital model.
For example, the present authors have shown that
 the robustness of $T_{\rm c}$ against impurities is inconsistent with the
 $s_{\pm}$-wave state. \cite{Onari_imp, Yamakawa_imp}
It has been shown that
 the broad resonant peak in the neutron scattering spectrum can be explained
 on the basis of the $s_{++}$-wave state rather than
 the $s_{\pm}$-wave state. \cite{Onari_neutron}
 Also, the absence of the coherence peak at $T_{\rm c}$
 can be explained in terms of both the $s_{++}$-wave and $s_{\pm}$-wave states. \cite{Yamakawa_nmr}
The theoretical study of the QPI signal in Fe-based superconductors ware performed
 by several theoretical groups
 in Refs.~\onlinecite{Sykora, Gao, Mazin_qpi, Zhang, Akbari, Das, Plamadeala}. 

By using the STM measurement, the information of the local density of states can be obtained. 
The QPI signal $Z ( \bm{q}, E )$ is given by the Fourier transformation of the tunneling conductance ratio
 $Z(\bm{r},E)=\frac{dI/dV(\bm{r},+V)}{dI/dV(\bm{r},-V)}$ derived from the STM measurement. 
 The QPI study played a crucial role to determine the pairing symmetry
 in cuprate superconductors. \cite{McElroy, Hanaguri_Cu, Maltseva}
In cuprate superconductors, the nodal $d_{x^2-y^2}$-wave SC state is realized. 
There are eight $\bm{k}$ points ($\bm{k}_i$ : $i = 1 \sim 8$) on the FS
 satisfying the relation $E = | \Delta ( \bm{k}_i ) |$ for $E < \Delta^{\rm max}$. 
It is called the octet model, 
 and the QPI signal $Z ( \bm{q}, E )$ with $\bm{q} = \bm{k}_i - \bm{k}_j$
 emerges due to the impurity scattering
 when $\Delta ( \bm{k}_i )$ and $\Delta ( \bm{k}_j )$ have the opposite sign, 
 while it disappears when $\Delta ( \bm{k}_i )$ and $\Delta ( \bm{k}_j )$ have the same sign. 
The disappearance of the QPI signal is called the ``extinction rule''. 
Furthermore, the experimental QPI peak is rapidly suppressed by applying a magnetic field. 
The extinction rule and the magnetic field dependence of the QPI 
 obtained in cuprate superconductors are well understood
 in terms of the octet model with $d_{x^2 - y^2}$-wave gap symmetry.

In Fe-based superconductors,
 many experimental \cite{Hanaguri, Hanaguri_arxiv, Allan, Hanke, Chi, Teague, Cai}
 and theoretical \cite{Sykora, Gao, Mazin_qpi, Zhang, Akbari, Das, Plamadeala}
 studies of the STM have been performed. 
Hanaguri {\it et al.} carried out the QPI experiments on Fe(Se,Te) single crystal
 and reported the appearance of a shape peak around $\bm{q}_2 = ( \pi, 0 )$, \cite{Hanaguri, Hanaguri_arxiv}
 which is caused by the impurity scattering between hole and electron FSs. 
By analogy with the extinction rule in the octet model, 
 the existence of the QPI peak around $\bm{q}_2$ 
 may indicate 
 that the gap functions on the hole and electron FSs have opposite sign, i.e., $s_{\pm}$-wave state.  
Although, many pioneering theoretical studies had been performed for Fe-based superconductors,
 some previous theoretical studies assumed over-simplified band structures. 
Furthermore, the QPI signal in the $s_{++}$-wave state had not been studied in detail in previous studies. 
Therefore, detailed theoretical study of the QPI based on a realistic five-orbital model
 in both the $s_{++}$-wave and $s_{\pm}$-wave states
 had been required. 

Hanaguri {\it et al.} showed that the intensity of the QPI peak around $\bm{q}_2$ is slightly suppressed
 by the magnetic field $B = 11$~T at $E = 1.0$~meV. \cite{Hanaguri}
However, the field-induced change of the QPI peak around $\bm{q}_2$ non-monotonically depends on $E$; 
 the peak intensity is slightly enhanced at $E = 0.5$~meV and $E = 1.9$~meV.
 [See Fig.~3S(I) in the Supplemental Material of Ref.~\onlinecite{Hanaguri}
 and Fig.~1(A) in Ref.~\onlinecite{Hanaguri_arxiv}.]
Therefore, in this paper, we discuss the field-induced change of the QPI for wide range of $E$
 in terms of both the $s_{++}$-wave and $s_{\pm}$-wave states. 

In this paper, we investigate the QPI in Fe-based superconductors
 on the bases of both the $s_{++}$-wave and $s_{\pm}$-wave states. 
In the cuprate superconductors with $d_{x^2 - y^2}$-wave SC state, 
 the QPI signal at $\bm{q} = \bm{k}_i - \bm{k}_j$ disappears
 when $\Delta ( \bm{k}_i )$ and $\Delta ( \bm{k}_j )$ have the same sign. 
However, such extinction rule
 does not hold in Fe-based superconductors with fully-gapped $s$-wave SC state, 
 since the resonance condition $E = | \Delta ( \bm{k}_i ) | = | \Delta ( \bm{k}_j ) |$
 is not satisfied under the experimental condition $E < \Delta^{\rm min}$
 regardless of the sign of the gap functions. 
We perform the detailed numerical study of the QPI signal based on the five-orbital model,
 and find that the experimentally observed QPI peak around $\bm{q}_2 = ( \pi, 0 )$
 appears in both the $s_{++}$-wave and $s_{\pm}$-wave states. 
Furthermore, we discuss the magnetic field dependence of the QPI by considering the Zeeman effect,
 and find that the field-induced change of the peak intensity around $\bm{q}_2$ can also be explained
 in terms of both the $s_{++}$-wave and $s_{\pm}$-wave states. 
In conclusion, it is difficult to distinguish between
the $s_{++}$-wave and $s_{\pm}$-wave states
 from the QPI experiments in Fe-based superconductors.

%%%%%%%%%%%%%%%%%%%%%%%%%%%%%%%%%%%%%%%%
%%%%%%%%%%%%%%%%%%%%%%%%%%%%%%%%%%%%%%%%

%%%%%%%%%%%%%%%%%%%%%%%%%%%%%%%%%%%%%%%%%%%%%%%%%%%%%%%%%%%%%%%%%%%%%%%%%%%%%%%%%%%%%%%%%%%%%%%%%%%%%%%%%%%%%%%%%%%%%%%%
\section{Formulation}
%%%%%%%%%%%%%%%%%%%%%%%%%%%%%%%%%%%%%%%%%%%%%%%%%%%%%%%%%%%%%%%%%%%%%%%%%%%%%%%%%%%%%%%%%%%%%%%%%%%%%%%%%%%%%%%%%%%%%%%%
\subsection{Quasiparticle Interference}
The tunneling conductance $dI / dV ( \bm{r}, V )$ at position $\bm{r}$ and voltage $V$
 is approximately proportional to the local density of states $\rho( \bm{r}, E)$ at energy $E = V$, 
namely, $dI / dV ( \bm{r}, V ) \propto \left| M ( \bm{r} ) \right|^2 \rho ( \bm{r}, E )$, 
where we set the unit of charge $e$ as one. 
$M ( \bm{r} )$ is the tunneling matrix element between the sample surface and the STM tip.
In the presence of the impurities, 
 we can drop the factor $M ( \bm{r} )$
 and obtain the information of the density of states by taking the ratio $Z ( \bm{r}, E )$
 between the conductance measured at $+V$ and $-V$ as follows: \cite{Maltseva, Sykora}
%%%%%%%%%%%%%%%%%%%%%%%%%%%%%%%%%%%%%%%%
\begin{eqnarray} \label{eq-zr}\
	Z ( \bm{r}, E )
& \equiv &
	\frac{ dI / dV ( \bm{r}, +V ) }
		 { dI / dV ( \bm{r}, -V ) } 
  =	\frac{ \rho ( \bm{r}, +E ) }
		 { \rho ( \bm{r}, -E ) }
\nonumber \\
& \approx &
	\frac{\rho^0 ( +E )}
		 {\rho^0 ( -E )}
	\left[
		1 + \frac{\delta \rho ( \bm{r}, +E )}{\rho^0 ( \bm{r}, +E )}
		  - \frac{\delta \rho ( \bm{r}, -E )}{\rho^0 ( \bm{r}, -E )}
	\right]
,
\end{eqnarray}
%%%%%%%%%%%%%%%%%%%%%%%%%%%%%%%%%%%%%%%%
 where $\rho^0 ( E )$ is the averaged density of states
 and $\delta \rho ( \bm{r}, E )$ describes the spatial modulation
 defined as $\delta \rho ( \bm{r}, E ) \equiv \rho ( \bm{r}, E ) - \rho^0 ( E )$. 
The Fourier transformed conductance ratio is called the QPI signal, which is given by
%%%%%%%%%%%%%%%%%%%%%%%%%%%%%%%%%%%%%%%%
\begin{eqnarray} \label{eq-zq}
	Z ( \bm{q}, E )
  =	\frac{ \rho^0 ( +E ) }{ \rho^0 ( -E ) }
% 	\left[
	&&\bigg[
		( 2 \pi )^2 \delta ( \bm{q} )
\nonumber \\
	&&
	+	\frac{ \delta \rho ( \bm{q}, +E ) }{ \rho^0 ( +E ) }
	  -	\frac{ \delta \rho ( \bm{q}, -E ) }{ \rho^0 ( -E ) }
% 	\right]
	\bigg]
,
\end{eqnarray}
%%%%%%%%%%%%%%%%%%%%%%%%%%%%%%%%%%%%%%%%
 where $\bm{q}$ is a scattering wave vector. 
When the system is uniform, $Z ( \bm{q}, E )$ is zero except for $\bm{q} = 0$. 
We can obtain the information on the SC gap symmetry
 since the momentum dependence of $Z ( \bm{q}, E )$ reflects the sign of the gap functions. 

%%%%%%%%%%%%%%%%%%%%%%%%%%%%%%%%%%%%%%%%%%%%%%%%%%%%%%%%%%%%%%%%%%%%%%%%%%%%%%%%%%%%%%%%%%%%%%%%%%%%%%%%%%%%%%%%%%%%%%%%
\subsection{Model Hamiltonian and Green Function}
The five-orbital tight-binding Hamiltonian is given by
\begin{eqnarray}
	\mathcal{H}^{0}
  =	\sum_{\bm{k}, l, l', \sigma}
		H_{\bm{k}, l,l'}^{0}
		c_{\bm{k}, l, \sigma}^{\dagger}
		c_{\bm{k}, l', \sigma}
\end{eqnarray}
 where $c_{\bm{k}, l, \sigma}^{\dagger}$ ($c_{\bm{k}, l, \sigma}$) is the creation (annihilation) operator
 of a Fe $3d$ electron with wave vector $\bm{k}$, orbital $l$ and spin $\sigma$. 
$\hat{H}_{\bm{k}}^{0}$ is given by the Fourier transformation of the hopping integrals
 introduced in Ref.~\onlinecite{Kuroki}. 
The energy dispersion $\epsilon_{\bm{k}, b}$ of band $b$ is obtained
 as a eigenvalue of $\hat{H}_{\bm{k}}^{0}$ by unitary transformation, 
\begin{eqnarray}
	\epsilon_{\bm{k}, b}
  =	\sum_{l, l'}
	U_{\bm{k}, l, b}^{*}
	H_{\bm{k}, l, l'}^{0}
	U_{\bm{k}, l', b}
,
\end{eqnarray}
 where $U_{\bm{k}, l, b}$ is an element of the unitary matrix obtained as the eigenvector. 
The obtained Fermi Surface is shown in Fig.~\ref{fig1}.  

Now we study the SC state. 
In a single-orbital model, the BCS Hamiltonian is simply given by 
\begin{eqnarray}
	\mathcal{H}
&=&	\sum_{\bm{k}}
	\hat{\Psi}_{\bm{k}}^{\dagger}
	\hat{H}_{\bm{k}}
	\hat{\Psi}_{\bm{k}}
,
\end{eqnarray}
 where 
\begin{eqnarray}
	\hat{\Psi}_{\bm{k}}^{\dagger}
\equiv
	\left(
		c_{ \bm{k}, \uparrow}^{\dagger}, 
		c_{-\bm{k}, \downarrow}^{\dagger}, 
		c_{-\bm{k}, \downarrow}, 
	  -	c_{ \bm{k}, \uparrow}
	\right)
.
\end{eqnarray}
Here, we define the Pauli matrices $\hat{\tau}_i$ and $\hat{\sigma}_i$
 which act in particle-hole space and spin space, respectively. 
For example, 
\begin{eqnarray}
	\hat{\tau}_1
&=&	\left(
	\begin{array}{cccc}
		 0 &  0 &  1 &  0 \\
		 0 &  0 &  0 &  1 \\
		 1 &  0 &  0 &  0 \\
		 0 &  1 &  0 &  0 
	\end{array}
	\right)
, \quad
	\hat{\tau}_3
  =	\left(
	\begin{array}{cccc}
		 1 &  0 &  0 &  0 \\
		 0 &  1 &  0 &  0 \\
		 0 &  0 & -1 &  0 \\
		 0 &  0 &  0 & -1 
	\end{array}
	\right)
,
\nonumber \\
	\hat{\sigma}_3
&=&	\left(
	\begin{array}{cccc}
		 1 &  0 &  0 &  0 \\
		 0 & -1 &  0 &  0 \\
		 0 &  0 &  1 &  0 \\
		 0 &  0 &  0 & -1 
	\end{array}
	\right)
.
\end{eqnarray}
Then, the Nambu Hamiltonian $\hat{H}_{\bm{k}}$ for a single-orbital model is given by
\begin{eqnarray}
	\hat{H}_{\bm{k}}
  =	  \epsilon_{\bm{k}} \hat{\tau}_3 
	- B \hat{\sigma}_{3}
	+ \Delta_{\bm{k}} \hat{\tau}_{1}
,
\end{eqnarray}
 where $B$ is the Zeeman splitting energy by the magnetic field
 and $\Delta_{\bm{k}}$ is the singlet gap function. 

In the five-orbital model, the Nambu Hamiltonian is written as follows: 
\begin{eqnarray}
	\hat{H}_{\bm{k}}
  =	\hat{H}_{\bm{k}}^{0} \hat{\tau}_{3}
  -	B \hat{E}_{5} \hat{\sigma}_{3}
  +	\hat{\Delta}^{\rm orb}_{\bm{k}} \hat{\tau}_{1}
,
\end{eqnarray}
 where $\hat{E}_5$ is the $5 \times 5$ unit matrix in the orbital space. 
In the case of the five-orbital model,
 the Nambu Hamiltonian is given by the $20 \times 20$ matrix form. 
$\hat{\Delta}^{\rm orb}_{\bm{k}}$ is $5 \times 5$ matrix form singlet gap function in the orbital space,
 and its matrix element is obtained by the unitary transformation
 of the band-basis gap function $\Delta_{\bm{k}, b}$ as 
\begin{eqnarray}
	\Delta^{\rm orb}_{\bm{k}, l, l'}
	\equiv
	\sum_{b} U_{\bm{k}, l, b} \Delta_{\bm{k}, b} U_{\bm{k}, l', b}^{*}
.
\end{eqnarray}
Then, the Green function in the clean limit is given by
\begin{eqnarray} \label{eq-g0}
	\hat{G}^{0}_{\bm{k}} ( \omega )
&=& \left( \omega - \hat{H}_{\bm{k}} \right)^{-1}
\nonumber \\
&=&	\left(
		\omega
	  -	\hat{H}_{\bm{k}}^{0} \hat{\tau}_{3}
	  +	B \hat{E}_{5} \hat{\sigma}_{3}
	  -	\hat{\Delta}^{\rm orb}_{\bm{k}} \hat{\tau}_{1}
	\right)^{-1}
, 
\end{eqnarray}
and the local density of states without randomness is given by 
\begin{eqnarray} \label{eq-dr0}
	\rho^{0} ( \omega )
= -	\frac{ 1 }{ \pi N }
	\sum_{\bm{k}}
	{\rm Im Tr}
	\hat{E}_{5}
	\frac{ \hat{\tau}_{0} + \hat{\tau}_{3} }{ 2 }
	\left.
	\hat{G}^{0}_{\bm{k}} ( \bar{\omega} )
	\right|_{\bar{\omega} = \omega + i \gamma}
, 
\end{eqnarray}
 where $\gamma$ is the quasiparticle damping rate. 

When we consider the impurity scattering, 
 the Green function is obtained by using the $T$-matrix approximation as follows: 
\begin{eqnarray}\label{eq-gkk}
	\hat{G}_{\bm{k}, \bm{k}'} ( \omega )
=
	\hat{G}^0_{\bm{k}} ( \omega ) \delta_{\bm{k}, \bm{k}'}
  +	\delta \hat{G}_{\bm{k}, \bm{k}'} ( \omega )
,
\end{eqnarray}
where 
\begin{eqnarray}\label{eq-dgkk}
	\delta \hat{G}_{\bm{k}, \bm{k}'} ( \omega )
\equiv
  	\hat{G}^0_{\bm{k}} ( \omega ) \hat{T}_{\bm{k}, \bm{k}'} ( \omega ) \hat{G}^0_{\bm{k}'} ( \omega ).
\end{eqnarray}
For a single impurity, the $T$-matrix is obtained by solving the following self-consistent equation, 
\begin{eqnarray}\label{eq-tkk}
	\hat{T}_{\bm{k}, \bm{k}'} ( \omega )
  =	\hat{I}_{\bm{k}, \bm{k}'}
  +	\frac{1}{N} \sum_{\bm{k}"}
	\hat{I}_{\bm{k}, \bm{k}"}
	\hat{G}^{0}_{\bm{k}"} ( \omega )
	\hat{T}_{\bm{k}", \bm{k}'} ( \omega )
, 
\end{eqnarray}
 where $\hat{I}_{\bm{k}, \bm{k'}}$ is the impurity potential of a single impurity. 
The modulation of the density of states induced by the impurity scattering
 is given by \cite{Maltseva, Sykora}
\begin{eqnarray}\label{eq-dr}
	\delta \rho ( \bm{q}, \omega )
= -	\frac{ n^{\rm imp} }{ \pi N }
	\sum_{\bm{k}}
	{\rm Im Tr}
	\hat{E}_{5}
	\frac{ \hat{\tau}_{0} + \hat{\tau}_{3} }{ 2 }
	\left.
	\delta \hat{G}_{\bm{k}, \bm{k+q}} ( \bar{\omega} )
	\right|_{ \bar{\omega} = \omega + i \gamma }
,
\end{eqnarray}
where $\hat{G}$ is represented in the orbital basis.
This treatment is exact for the case of low impurity concentration $n^{\rm imp} \ll 1$. 

In this paper, we consider the non-magnetic impurity
 since the QPI due to the magnetic impurity scattering is subdominant for $B = 0$. \cite{Sykora} 
According to the band calculations, 
 the impurity potential in Fe-based superconductors is screened and well-localized. \cite{Nakamura}
That is, the impurity scattering matrix in the orbital space is $\bm{k}$-independent. 
When the Fe-site substitution is considered, the impurity potential is given as 
\begin{eqnarray}\label{eq:iimp}
	\hat{I}^{\rm imp}
  =	I^{\rm imp} \hat{E}_{5} \hat{\tau}_{3}
,
\end{eqnarray}
and then, the $T$-matrix becomes $\bm{k}$-independent and it is simply given by
\begin{eqnarray}\label{eq:tmat}
	\hat{T} ( \omega )
  =	\left(
		1 - \hat{I}^{\rm imp} \hat{g}^{0} ( \omega )
	\right)^{-1}
	\hat{I}^{\rm imp}
,
\end{eqnarray}
where $\hat{g}^{0} ( \omega ) \equiv \frac{1}{N} \sum_{\bm{k}} \hat{G}^{0}_{\bm{k}} ( \omega )$
 is the local Green function in the $20 \times 20$ matrix form.

%%%%%%%%%%%%%%%%%%%%%%%%%%%%%%%%%%%%%%%%%%%%%%%%%%%%%%%%%%%%%%%%%%%%%%%%%%%%%%%%%%%%%%%%%%%%%%%%%%%%%%%%%%%%%%%%%%%%%%%%
\section{Result} \label{sec-result}
\subsection{Simple Analytical Calculation} \label{sec-ana}
In this section, we analytically show that the extinction rule, 
 which tells that
 the non-magnetic impurity scattering between FSs with same sign gap functions
 does not contribute to the QPI, 
 does not hold in fully gapped $s$-wave SC state. 

Here, we verify the case with the particle-hole symmetry $\rho^{0} ( +E ) = \rho^{0} ( -E )$. 
Then, $Z ( \bm{q}, E )$ in Eq.~(\ref{eq-zq}) is simplified as 
\begin{eqnarray}
	Z ( \bm{q} \ne \bm{0}, E )
  =	\frac{ 2 \delta \rho^{\rm odd} ( \bm{q}, E ) }
		 { \rho^0 ( E ) }
,
\end{eqnarray}
where
\begin{eqnarray}
	\delta \rho^{\rm odd} ( \bm{q}, E )
\equiv
	\frac{\delta \rho ( \bm{q}, +E )
	     -\delta \rho ( \bm{q}, -E )}{ 2 }
.
\end{eqnarray}

When we consider the scattering due to non-magnetic impurities with a weak scalar potential $I^{\rm imp}$, 
 the $T$-matrix is given by $\hat{T} \approx I^{\rm imp} \hat{E}_{5} \hat{\tau}_{3}$. 
From Eq.~(\ref{eq-dr}), the modulation of the density of states 
 for $\hat{T} \approx I^{\rm imp} \hat{E}_5 \hat{\tau}_3$
 is given by
\begin{eqnarray} \label{eq-drodd}
&&	\delta \rho^{\rm odd} ( \bm{q}, E ) 
=
  -	\frac{n^{\rm imp}}{ \pi N }
	\sum_{\bm{k}}
	{\rm Im Tr}
	\hat{E}_{5}
	\frac{ \hat{\tau}_{3} }{ 2 }
	\delta \hat{G}_{\bm{k}, \bm{k} + \bm{q} } ( \bar{E} )
	\Big|_{ \bar{E} = E + i \gamma }
\nonumber \\
& \approx &
  -	\frac{ n^{\rm imp} I^{\rm imp} }{ 2 \pi N }
	\sum_{\bm{k}, b, b'}
	{\rm Im}
	\left.
	\frac{ \bar{E}^2
		 + \epsilon_{\bm{k}, b} \epsilon_{\bm{k} + \bm{q}, b'}
		 - \Delta_{\bm{k}, b} \Delta_{\bm{k} + \bm{q}, b'}}
		 { ( \bar{E}^2 - E_{\bm{k}, b}^2 )
		   ( \bar{E}^2 - E_{\bm{k} + \bm{q}, b'}^2 ) }
	\right|_{ \bar{E} = E + i \gamma }
\nonumber \\
 &&	\times
	\left|
	\sum_{l}
		U_{\bm{k}, l, b}
		U_{\bm{k} + \bm{q}, l, b'}^{*}
	\right|^2
.
\end{eqnarray}
In the last line, we utilized the functional form of the Green function in the band-diagonal basis,
 and $E_{\bm{k}, b}^2 \equiv \epsilon_{\bm{k}, b}^2 + \Delta_{\bm{k}, b}^2$
 is the energy of a quasiparticle in band $b$. 
In Eq.~(\ref{eq-drodd}), the main contribution originates from the case
 that both $\bm{k}$ and $\bm{k} + \bm{q}$ are on FSs
 ($\epsilon_{ \bm{k}, b } = \epsilon_{ \bm{k} + \bm{q}, b' } = 0$). 
In this case, the contribution is simplified as
\begin{eqnarray} \label{eq-drodd-d}
	\delta \rho^{\rm odd} ( \bm{q}, E ) 
\propto
  -	\!\!\!\!
	\sum_{\bm{k}, \bm{k} + \bm{q} \in {\rm FS}}
	\!\!\!\!
	{\rm Im}
	\left.
	\frac{ \bar{E}^2
		 - \Delta_{\bm{k}} \Delta_{\bm{k} + \bm{q}}}
		 { ( \bar{E}^2 - \Delta_{\bm{k}}^2 )
		   ( \bar{E}^2 - \Delta_{\bm{k} + \bm{q}}^2 ) }
	\right|_{ \bar{E} = E + i \gamma }
.
\end{eqnarray}

In cuprate superconductors, the nodal $d_{x^2-y^2}$-wave SC state is realized. 
Under the experimental condition $E < \Delta^{\rm max}$,
 only the eight $\bm{k}$ points ($\bm{k}_i$ : $i = 1 \sim 8$)
 satisfy the relation $E = | \Delta_{\bm{k}_i} |$. 
It is called the octet model,\cite{McElroy, Hanaguri_Cu, Maltseva}
 and $\bm{k}_1 \sim \bm{k}_8$ are shown in Fig.~\ref{figa1}(a). 
$\delta \rho^{\rm odd} ( \bm{q}, E )$ can be very large for $\bm{q} = \bm{k}_i - \bm{k}_j$
 since the denominator in Eq.~(\ref {eq-drodd-d}) is almost zero for $\bm{k} \approx \bm{k}_i$.  
On the other hand, the numerator is sensitive to the sign of the gap functions: 
the numerator has finite value $2 E^2$
 when the gap functions $\Delta_{\bm{k}}$ and $\Delta_{\bm{k} + \bm{q}}$ have opposite sign, 
 but it becomes zero for the same sign case. 
Therefore, the QPI peak disappears when the gap functions at $\bm{k}$ and $\bm{k} + \bm{q}$ have the same sign, 
 which is called the extinction rule. 

In contrast, such extinction rule does not hold
 in the fully gapped $s$-wave SC state realized in Fe-based superconductors, 
 under the experimental condition $E < \Delta^{\rm min}$. 
We focus on the QPI peak around $\bm{q}_2 = ( \pi, 0)$
 which corresponds to the inter-band scattering between the hole and electron FSs. 
Using the gap functions on hole FS $\Delta_{\rm h}$ and electron FS $\Delta_{\rm e}$,
 $\delta \rho^{\rm odd} ( \bm{q}_2, E )$ is given by
\begin{eqnarray} \label{eq-drodd-s}
	\delta \rho^{\rm odd} ( \bm{q}_2, E ) 
\propto
  -	{\rm Im}
	\left.
	\frac{ \bar{E}^2
		 - \Delta_{\rm h} \Delta_{\rm e}}
		 { ( \bar{E}^2 - \Delta_{\rm h}^2 )
		   ( \bar{E}^2 - \Delta_{\rm e}^2 ) }
	\right|_{ \bar{E} = E + i \gamma }
.
\end{eqnarray}
In the case of $E < | \Delta_{\rm h, e} |$,
 both the numerator and denominator have finite value
 regardless of the signs of $\Delta_{\rm h}$ and $\Delta_{\rm e}$. 
That is, $\delta \rho^{\rm odd} ( \bm{q}_2, E )$ is finite even for $\Delta_{\rm h} \Delta_{\rm e} > 0$.
Therefore, the extinction rule does not hold in Fe-based superconductors, 
 and the QPI peak around $\bm{q}_2$ is expected to
 appear in both the $s_{++}$-wave and $s_{\pm}$-wave states. 
We will numerically verify the violation of the extinction rule for the $\bm{q}_2$ signal
 by analyzing the five-orbital model in later sections. 
 
As shown in Fig.~\ref{fig1}, 
 the other QPI signal can arise around $\bm{q}_1 = ( 0, 0 )$ due to intra-band scattering,
 and around $\bm{q}_3 = ( \pi, \pi )$ due to inter-band scattering between hole-FSs or electron-FSs. 
Experimentally, the QPI peak around $\bm{q}_3$ is enhanced by the external magnetic field. 
However, both the QPI peaks around $\bm{q}_1$ and $\bm{q}_3$ are caused
 by the scattering between hole-pockets and between electron-pockets. 
These QPI peaks are not useful for the purpose of distinguishing
 between the $s_{++}$-wave and $s_{\pm}$-wave states.

%%%%%%%%%%%%%%%%%%%%%%%%%%%%%%%%%%%%%%%%%%%%%%%%%%%%%%%%%%%%%%%%%%%%%%%%%%%%%%%%%%%%%%%%%%%%%%%%%%%%%%%%%%%%%%%%%%%%%%%%
\subsection{QPI for the Weak Impurity Potential Case} \label{sec-born}
%%%%%%%%%%%%%%%%%%%%%%%%%%%%%%%%%%%%%%%%%%%%%%%%%%%%%%%%%%%%%%%%%%%%%%%%%%%%%%%%%%%%%%%%%%%%%%%%%%%%%%%%%%%%%%%%%%%%%%%%
In this and subsequent sections, we numerically calculate the QPI signal using Eq.~(\ref{eq-dr}). 
Here, we discuss the QPI 
 due to a weak impurity potential $I^{\rm imp} = 0.1$~eV, 
 and show that the QPI peak around $\bm{q}_2$ is actually obtained
 in both the $s_{++}$-wave and $s_{\pm}$-wave states for various parameters. 
Hereafter, we set $\Delta_0 = 0.02$~eV, $n^{\rm imp} = 0.01$, $\gamma = \Delta_0 / 4$ and $N = 256 \times 256$.
We confirmed that the obtained results do not change qualitatively for $\gamma = \Delta_0 / 8$.

%%%%%%%%%%%%%%%%%%%%%%%%%%%%%%%%%%%%%%%%
\begin{figure}[htb]
	\includegraphics[width=1.0\linewidth]{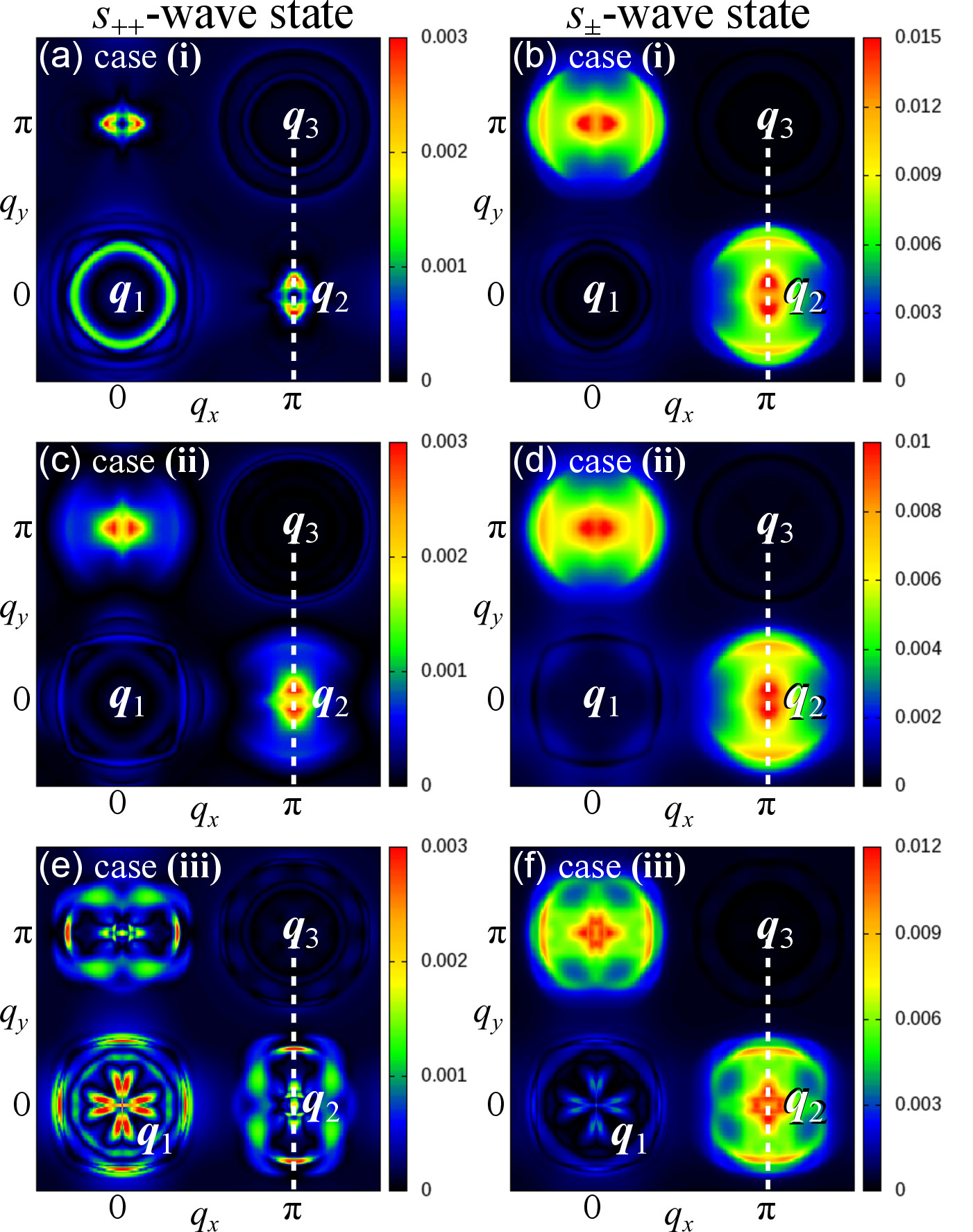}
	\caption{(Color online)
		Intensity map of the QPI $| Z ( \bm{q}, E ) |$ under zero field $B = 0$ at $E = \Delta_{0} / 2$
		 due to the non-magnetic impurity scattering with potential $I^{\rm imp} = 0.1$ eV. 
		The left and right panels show the results
		 in the $s_{++}$-wave and $s_{\pm}$-wave states, respectively. 
		(a),(b) Case {\bf (i)}: Isotropic single-gap case with $\Delta_{\rm h} = \pm \Delta_{\rm e} = \Delta_{0}$.
		(c),(d) Case {\bf (ii)}: Isotropic two-gap case
		 with $\Delta_{\rm h} = 2 \Delta_{0}$ and $\Delta_{\rm e} = \pm \Delta_{0}$.
		(e),(f) Case {\bf (iii)}: Strongly anisotropic gap case with $\Delta_{\rm h} = \Delta_{0}$,
		 $\Delta_{\rm e} = \pm ( 1 + \cos 2 \theta ) \Delta_{0}$ around $\bm{k} = ( \pi, 0 )$
		 and $\Delta_{\rm e} = \pm ( 1 - \cos 2 \theta ) \Delta_{0}$ around $\bm{k} = ( 0, \pi )$. 
		The vertical broken lines represent the path of the linecuts in Fig.~\ref{fig3}. 
		}
	\label{fig2}
\end{figure}
%%%%%%%%%%%%%%%%%%%%%%%%%%%%%%%%%%%%%%%%

Figure~\ref{fig2} shows the intensity map of the QPI, $| Z ( \bm{q}, E ) |$, at zero field. 
First, we discuss the {\bf (i)} isotropic single-gap case with $| \Delta_{\rm h} | = | \Delta_{\rm e} | = \Delta_{0}$: 
Figures~\ref{fig2}(a) and (b) show the results obtained
 in the $s_{++}$-wave and $s_{\pm}$-wave states, respectively. 
Considering the experimental condition in Ref.~\onlinecite{Hanaguri},
 we set $ E = \Delta_{0} / 2$. 
In the $s_{++}$-wave state (a),
 the sharp QPI peak around $\bm{q}_2$ clearly appears as expected from Eq.~(\ref{eq-drodd-s}). 
Therefore, the extinction rule does not hold in Fe-based superconductors. 
In the $s_{\pm}$-wave state (b),
 the strong QPI peak accompanied by the large halo structure is obtained around $\bm{q}_2$. 
That is, it is difficult to distinguish between the $s_{++}$-wave
 and $s_{\pm}$-wave states by the presence or absence of the QPI peak around $\bm{q}_2$. 

In reality, $| \Delta_{\rm h} |$ and $| \Delta_{\rm e} |$ are different in usual Fe-based superconductors. 
For example, $|\Delta^{\rm max} / \Delta^{\rm min}| \sim 2$ is reported
 in electron- and hole-doped BaFe$_2$As$_2$. \cite{Williams, Nakayama}
In the Fe(Se,Te) sample used in the QPI experiments, \cite{Hanaguri}
 the relations $\Delta_{\rm min} \sim 1$~meV and $\Delta_{\rm max} = 2 \sim 4$ are expected
 from the tunneling conductance measurement.
Therefore, we show the results for the {\bf (ii)} isotropic two-gap case
 with $| \Delta_{\rm h} | = 2 \Delta_{0}$ and $| \Delta_{\rm e} | = \Delta_{0}$ 
 in Figs.~\ref{fig2}(c) and (d). 
In this case, there is no large difference
 from the single-gap case shown in Figs.~\ref{fig2}(a) and (b). 
Similar results are obtained when $| \Delta_{\rm h} | = \Delta_{0}$ and $| \Delta_{\rm e} | = 2 \Delta_{0}$. 

In Figs.~\ref{fig2}(e) and (f), we also show the {\bf (iii)} strongly anisotropic gap case
 with $| \Delta_{\rm h} | = \Delta_{0}$ and $ | \Delta_{\rm e} | = ( 1 \pm \cos 2 \theta ) \Delta_{0} $. 
Anisotropic-gap functions are reported on a hole-FS in heavily K-doped BaFe$_2$As$_2$ \cite{Ota}
 and on the electron FSs in some Fe(Se,Te) systems. \cite{Zeng, Song}
In this case, the peak around $\bm{q}_2$ exists
 and its shape in the $s_{++}$-wave state becomes similar to the one in the $s_{\pm}$-wave state. 
Therefore, it is difficult to distinguish between
 the $s_{++}$-wave and $s_{\pm}$-wave states
 from the existence of the QPI signal around $\bm{q}_2$.

%%%%%%%%%%%%%%%%%%%%%%%%%%%%%%%%%%%%%%%%%%%%%%%%%%%%%%%%%%%%%%%%%%%%%%%%%%%%%%%%%%%%%%%%%%%%%%%%%%%%%%%%%%%%%%%%%%%%%%%%

Experimentally, the QPI peak intensity around $\bm{q}_2$ is slightly suppressed
 by the magnetic field $B$ for $E = 1.0$~meV~$\le \Delta^{\rm min}$. \cite{Hanaguri, Hanaguri_arxiv}
Here, we discuss $B-$ and $E-$ dependencies of $| Z ( \bm{q}, E ) |$ in detail, 
 and show that the experimental suppression of the $\bm{q}_2$ peak for $E \sim \Delta^{\rm min}$ can be explained
 in both the $s_{++}$-wave and $s_{\pm}$-wave states. 
Previously, two kinds of the field-induced suppression effects have been discussed
 by Coleman {\it et al.}: \cite{Sykora, Maltseva}
(A) Impurities are masked by vortices under the magnetic field, and then the impurity scattering rate is reduced. 
Also, (B) the Zeeman effect changes the electronic state and modifies the impurity scattering. 
The former mechanism would suppress the QPI intensity around $\bm{q}_2$ regardless of the sign of the gap functions. 
However, in the QPI experiments for Fe(Se,Te) in Ref.~\onlinecite{Hanaguri},
 it was reported that the effect (B) would be dominant,
 since the field-induced changes are almost spatially uniform. 
Therefore, in this paper, we study only the effect (B). 

%%%%%%%%%%%%%%%%%%%%%%%%%%%%%%%%%%%%%%%%q
\begin{figure}[tb]
	\includegraphics[width=0.45\linewidth]{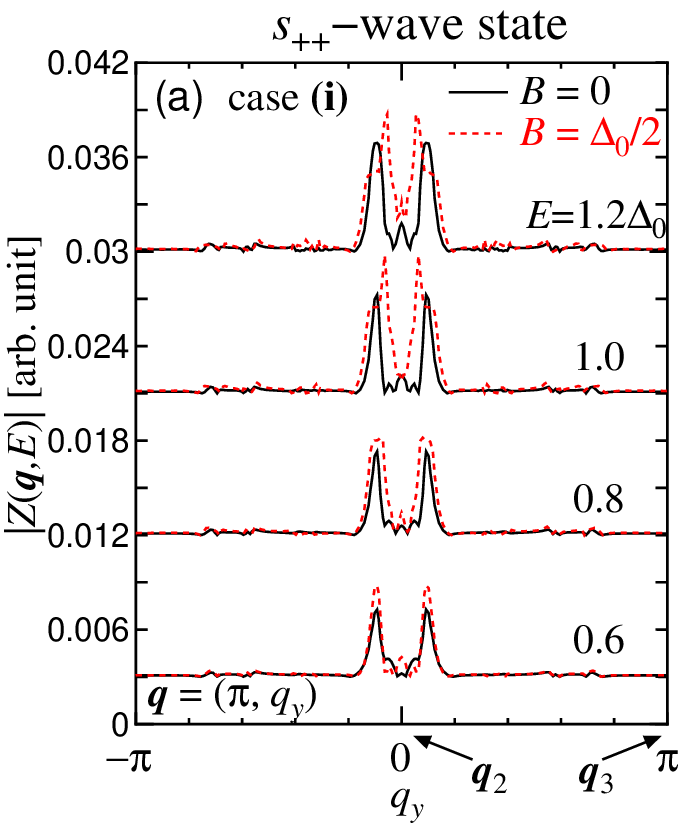}
	\includegraphics[width=0.45\linewidth]{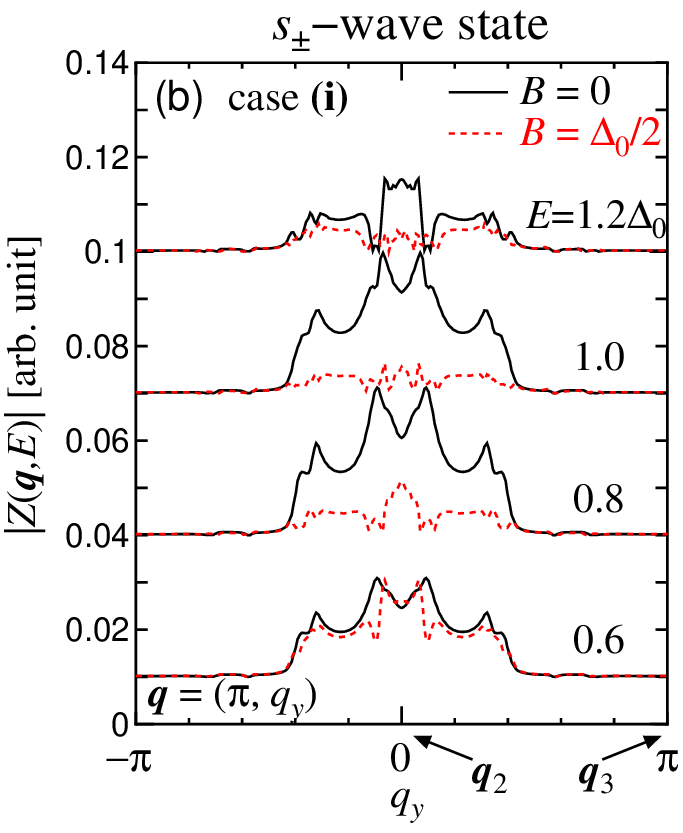}
	\includegraphics[width=0.45\linewidth]{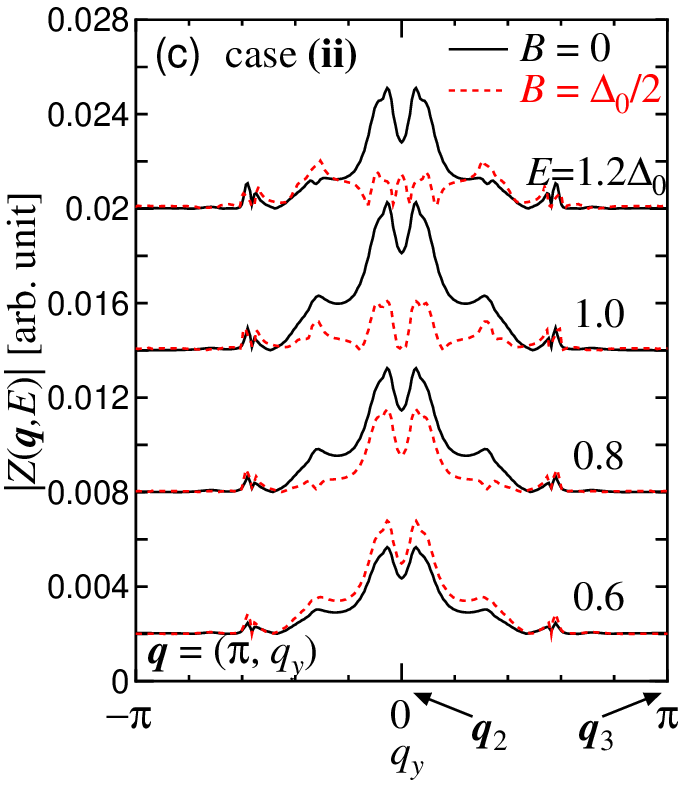}
	\includegraphics[width=0.45\linewidth]{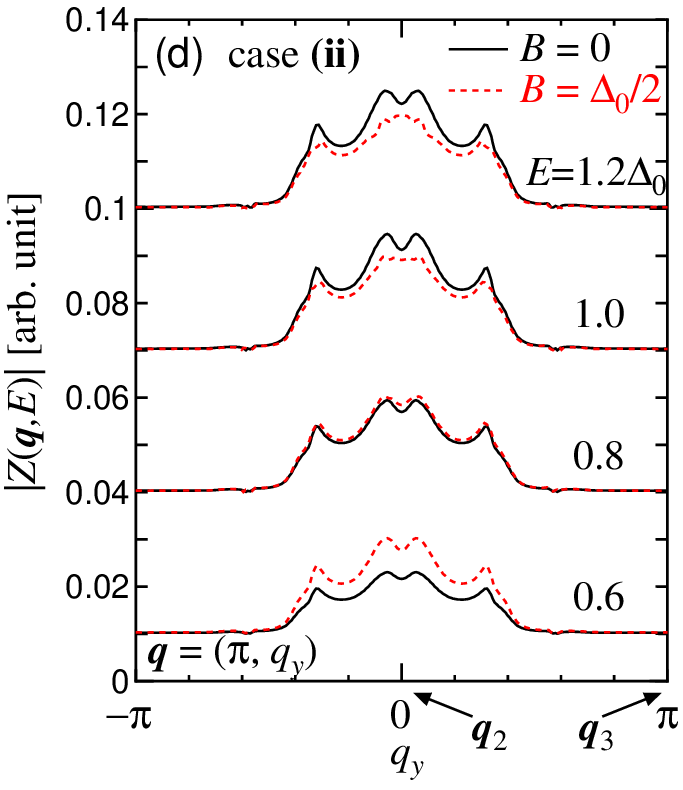}
	\includegraphics[width=0.45\linewidth]{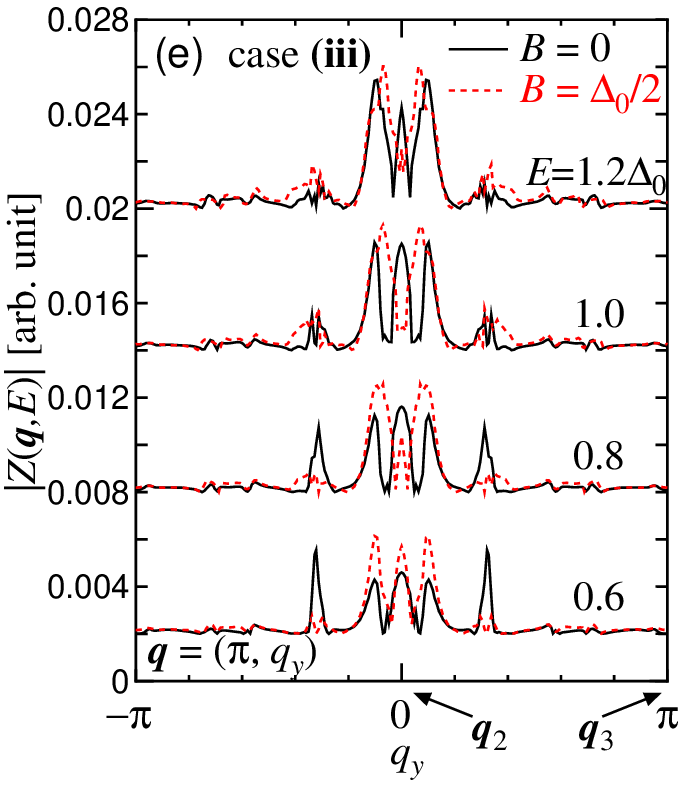}
	\includegraphics[width=0.45\linewidth]{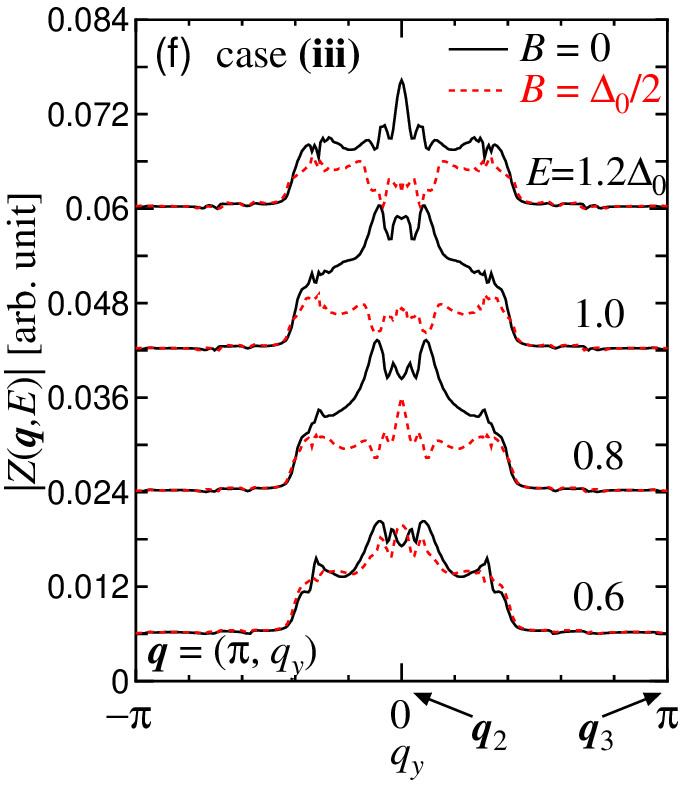}
	\caption{(Color online)
		Linecuts from $| Z ( \bm{q}, E ) |$ map for $I^{\rm imp} = 0.1$~eV. 
		The solid and dotted lines represent $| Z ( \bm{q}, E ) |_{B = 0}$
		 and $| Z ( \bm{q}, E ) |_{B = \Delta_{0} / 2}$, respectively. 
		The left and right panels show the results
		 in the $s_{++}$-wave and $s_{\pm}$-wave states, respectively. 
		The path is shown in Fig.~\ref{fig2}. 
		(a),(b) Case {\bf (i)}: Isotropic single-gap case with $\Delta_{\rm h} = \pm \Delta_{\rm e} = \Delta_{0}$. 
		(c),(d) Case {\bf (ii)}: Isotropic two-gap case
		 with $\Delta_{\rm h} = 2 \Delta_{0}$ and $\Delta_{\rm e} = \pm \Delta_{0}$.
		(e),(f) Case {\bf (iii)}: Strongly anisotropic gap case with $\Delta_{\rm h} = \Delta_{0}$,
		 $\Delta_{\rm e} = \pm ( 1 + \cos 2 \theta ) \Delta_{0}$ around $\bm{k} = ( \pi, 0 )$
		 and $\Delta_{\rm e} = \pm ( 1 - \cos 2 \theta ) \Delta_{0}$ around $\bm{k} = ( 0, \pi )$. 
		The curves in all the figures are vertically shifted to make them visible.
	}
	\label{fig3}
\end{figure}
%%%%%%%%%%%%%%%%%%%%%%%%%%%%%%%%%%%%%%%%

Figures~\ref{fig3}(a) and (b) show the $| Z ( \bm{q}, E ) |$
 in the single-gap case with $| \Delta_{\rm h} | = | \Delta_{\rm e} | = \Delta_{0}$ [case {\bf (i)}]
 from $\bm{q} = ( \pi, -\pi )$ to $( \pi, \pi )$. 
The path is shown in Fig.~\ref{fig2} by the vertical dashed lines. 
The solid and dotted lines represent the results for $B = 0$ and $B = \Delta_0 / 2$, respectively. 
In the (a) $s_{++}$-wave state, the QPI peak around $\bm{q}_2$ is not sensitive to $B$ and $E$. 
On the other hand, 
in the (b) $s_{\pm}$-wave state, the $\bm{q}_2$ peak is drastically suppressed by $B$.
Figures~\ref{fig3}(c) and (d) show the results obtained for the two-gap case [case {\bf (ii)}]. 
In this case, the QPI peak around $\bm{q}_2$
 is suppressed by $B$ for $E \sim \Delta_{0}$ in both the $s_{++}$-wave and $s_{\pm}$-wave states. 
However, the field suppression of the QPI peaks is much larger in the $s_{++}$-wave state. 
Figures~\ref{fig3}(e) and (f) show the results for the strongly anisotropic gap case [case {\bf (iii)}]. 
In this case, $| Z ( \bm{q}, E ) |$ in the $s_{++}$-wave state shows very complex $B$ dependence.

In summary, in the $s_{\pm}$-wave state,
 the QPI peak around $\bm{q}_2$ is clearly suppressed in all cases {\bf (i)}-{\bf (iii)}. 
In the $s_{++}$-wave state, this peak intensity is also suppressed in the two-gap case {\bf (ii)}.
Therefore, the field-induced suppression of the QPI around $\bm{q}_2$ can be explained
 in terms of both the $s_{++}$-wave and $s_{\pm}$-wave states. 
Experimentally, the SC gaps are fully opened in the Fe(Se,Te) sample used for the QPI experiments,
 and relation $\Delta^{\rm max} \gg \Delta^{\rm min} \sim 1$~meV is expected,
 since the estimated value of $2 \Delta^{\rm min} / T_{\rm c} < 2$ is much smaller than the BCS value $3.53$.
In addition, the tunneling conductance has the sharp gap edge peak at $V \approx 1.7$~mV and an additional peak at about $4$~mV. 
If the latter peak arises from the SC gap, $\Delta^{\rm max} / \Delta^{\rm min} \ge 2$ is expected. 
Therefore, the isotropic two-gap case with $| \Delta_{\rm h} | = 2 | \Delta_{\rm e} | $ [case {\bf (ii)}]
 would correspond to Fe(Se,Te).

%%%%%%%%%%%%%%%%%%%%%%%%%%%%%%%%%%%%%%%%%%%%%%%%%%%%%%%%%%%%%%%%%%%%%%%%%%%%%%%%%%%%%%%%%%%%%%%%%%%%%%%%%%%%%%%%%%%%%%%%
\subsection{QPI for the Strong Impurity Potential Case} \label{sec-uni}
In this section, we consider the QPI due to a strong impurity potential $| I^{\rm imp} | = 1$~eV,
 which corresponds to Fe-site substitution. 
Since the residual resistivity takes the maximum for $I^{\rm imp} \sim +1$~eV,
 $I^{\rm imp} = + 1$~eV corresponds to the unitary limit
 in Fe-based superconductors. \cite{Onari_imp, Yamakawa_imp}
Here, we show the result only for the isotropic two-gap case
 with $| \Delta_{\rm h} | = 2 \Delta_{0}$ and $| \Delta_{\rm e} | = \Delta_{0}$
 [case {\bf (ii)}]. 

\begin{figure}[tb]
	\includegraphics[width=1.0\linewidth]{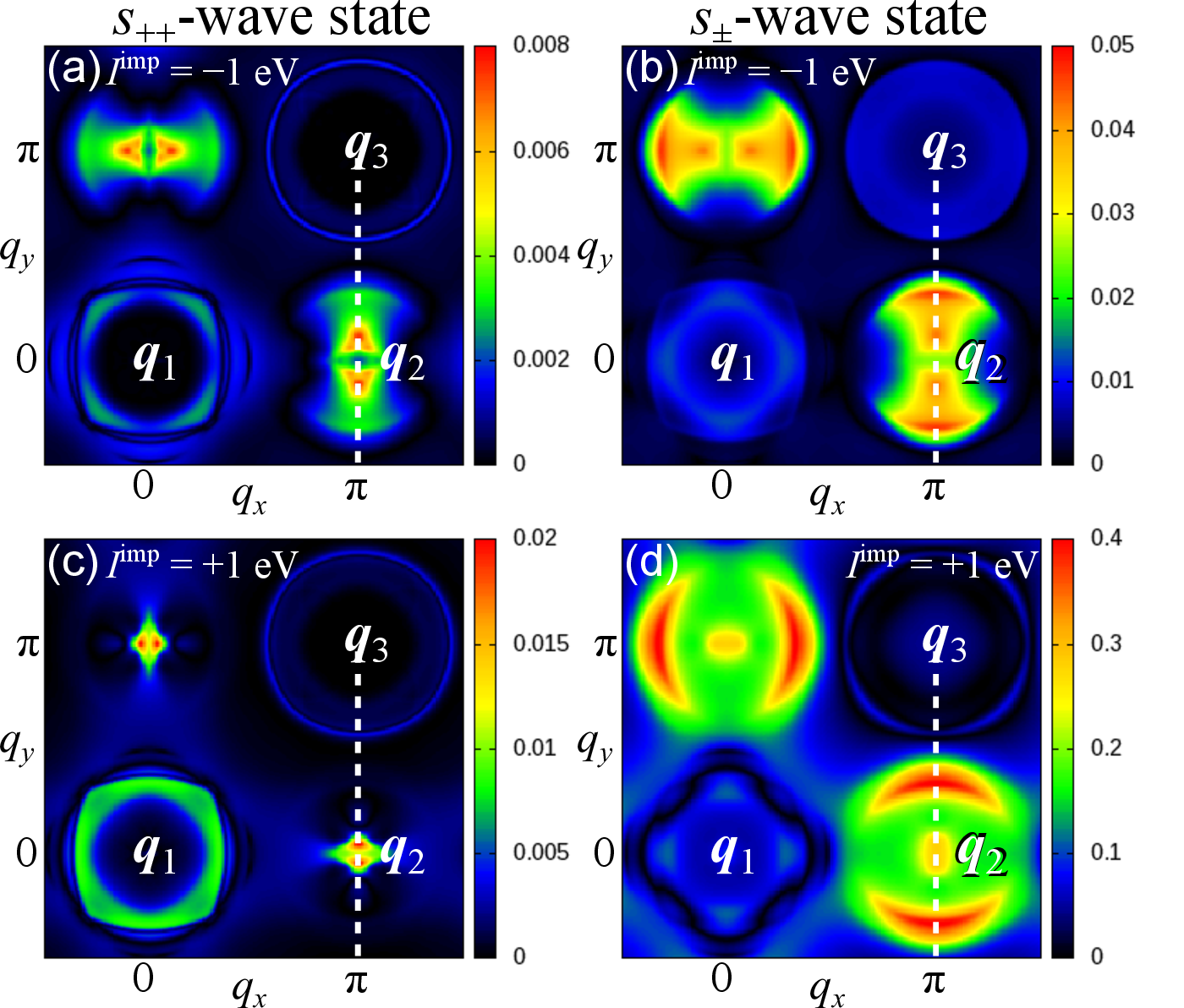}
	\caption{(Color online)
		Intensity map of $| Z ( \bm{q}, E ) |$ at $E = \Delta_{0} / 2$ and $B = 0$
		 in the isotropic two-gap case
		 with $\Delta_{\rm h} = 2 \Delta_{0}$ and $\Delta_{\rm e} = \pm \Delta_{0}$ [case {\bf (ii)}]. 
		The left and right panels show the results
		 in the $s_{++}$-wave and $s_{\pm}$-wave states, respectively. 
		(a),(b) $I^{\rm imp} = -1$~eV.
		(c),(d) $I^{\rm imp} = +1$~eV. 
		The vertical broken lines represent the path of the linecuts in Fig.~\ref{fig5}. 
	}
	\label{fig4}
\end{figure}

Figures~\ref{fig4}(a) and (b) show the $| Z ( \bm{q}, E ) |$ map for $I^{\rm imp} = -1$~eV
 in the case of the $s_{++}$-wave and $s_{\pm}$-wave states, respectively. 
Also, Figs.~\ref{fig4}(c) and (d) show the ones for $I^{\rm imp} = +1$~eV. 
We set $E = \Delta_{0} / 2$ and $B = 0$. 
The obtained QPI map is qualitatively similar to the ones in the weak potential case
% in the Born regime
 shown in Fig.~\ref{fig2}, 
 and the QPI peak around $\bm{q}_2$ appears in both the $s_{++}$-wave and $s_{\pm}$-wave states. 
Therefore, the extinction rule does not hold in Fe-based superconductors
 regardless of the magnitude of the impurity potential. 

%%%%%%%%%%%%%%%%%%%%%%%%%%%%%%%%%%%%%%%%
%%%%%%%%%%%%%%%%%%%%%%%%%%%%%%%%%%%%%%%%
%%%%%%%%%%%%%%%%%%%%%%%%%%%%%%%%%%%%%%%%
\begin{figure}[tb]
	\includegraphics[width=0.45\linewidth]{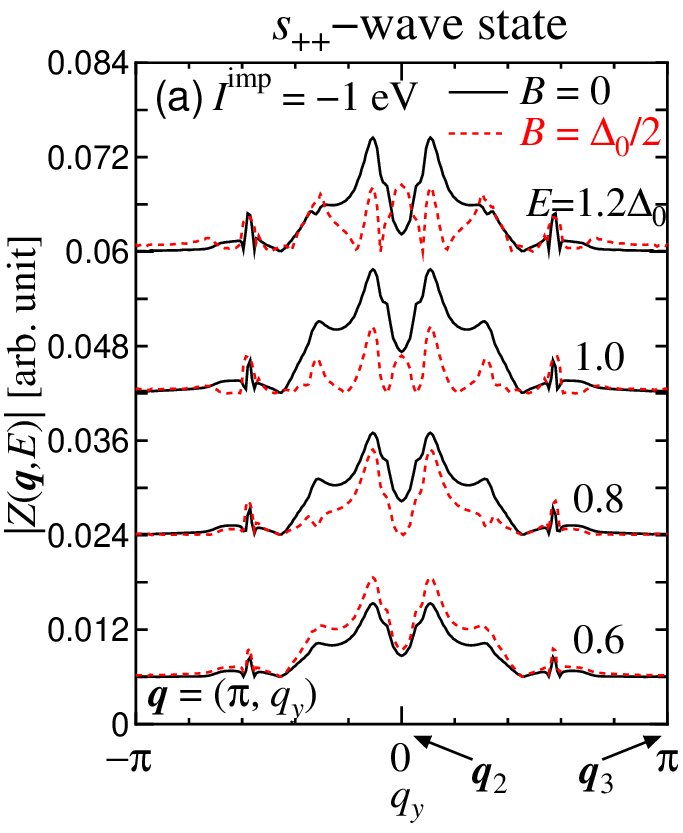}
	\includegraphics[width=0.45\linewidth]{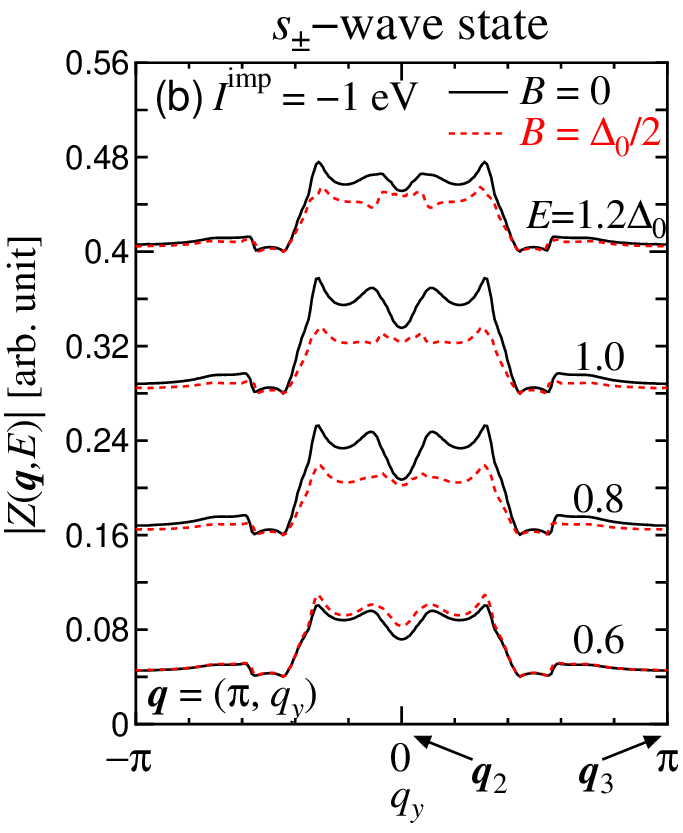}
	\includegraphics[width=0.45\linewidth]{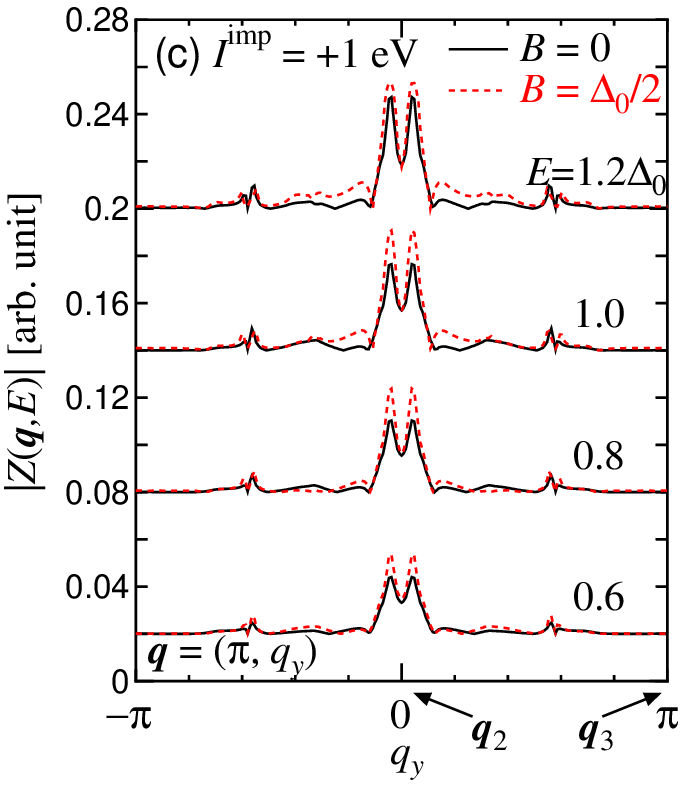}
	\includegraphics[width=0.45\linewidth]{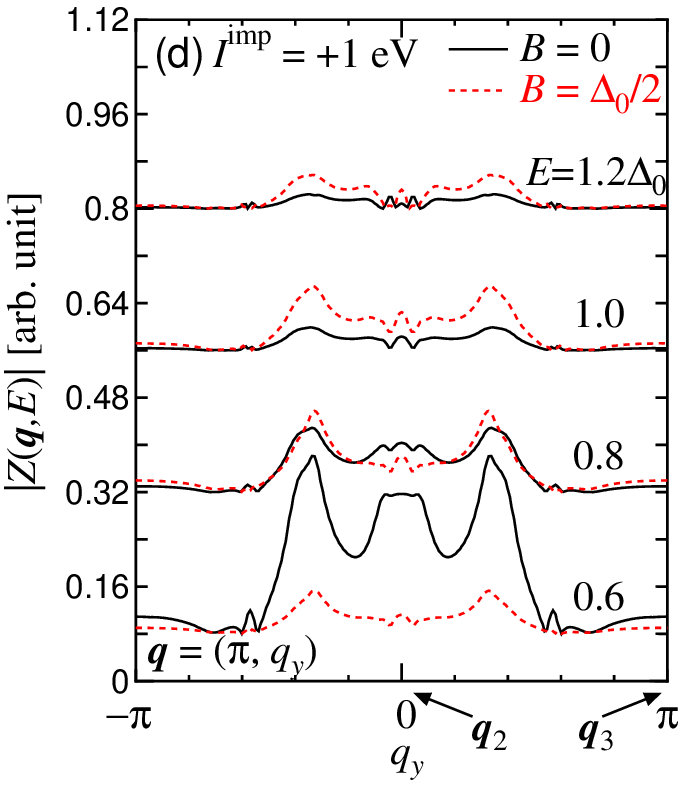}
	\caption{(Color online)
		Linecut from $| Z ( \bm{q}, E ) |_{B = 0}$ (solid lines)
		 and $| Z ( \bm{q}, E ) |_{B = \Delta_0 / 2}$ (dotted lines) 
		 maps in the isotropic two-gap case
		with $\Delta_{\rm h} = 2 \Delta_{0}$ and $\Delta_{\rm e} = \pm \Delta_{0}$ [case {\bf (ii)}]. 
		The path is shown in Fig.~\ref{fig4}. 
		The left and right panels show the results
		 in the $s_{++}$-wave and $s_{\pm}$-wave states, respectively. 
		(a),(b) $I^{\rm imp} = - 1$ eV.
		(c),(d) $I^{\rm imp} = + 1$ eV.
		The curves in all the figures are vertically shifted to make them visible. 
	}
	\label{fig5}
\end{figure}
%%%%%%%%%%%%%%%%%%%%%%%%%%%%%%%%%%%%%%%%

Figures~\ref{fig5}(a) and (b) show $| Z ( \bm{q}, E ) |$
 from $\bm{q} = ( \pi, - \pi )$ to $( \pi, \pi )$ for $I^{\rm imp} = -1$~eV. 
The solid and dotted lines represent the results for $B = 0$ and $\Delta_0 / 2$, respectively. 
For $E \sim \Delta_{0}$, the QPI peak around $\bm{q}_2$ is suppressed by the magnetic field
 in both the $s_{++}$-wave and $s_{\pm}$-wave states. 

Figures~\ref{fig5}(c) and (d) show the results for $I^{\rm imp} = +1$~eV. 
In the $s_{++}$-wave state,
 the QPI peak around $\bm{q}_2$ is insensitive to $B$ and $E$. 
On the other hand, in the $s_{\pm}$-wave state,
 the QPI signal shows very strong $E$ dependence,
 and the QPI intensity becomes very small for $E \ge 0.8 \Delta_0$ even for $B = 0$. 
However, such behaviors have not been observed experimentally. 
As results, in both the $s_{++}$-wave and $s_{\pm}$-wave states,
 the obtained results for $I^{\rm imp} = +1$~eV are inconsistent
 with experiments. \cite{Hanaguri, Hanaguri_arxiv}
Therefore, impurities with weak potential will be responsible for the QPI signal in Fe(Se,Te). 

In the above discussion, we have ignored the change of $T_{\rm c}$ due to the impurity scattering. 
We have shown that the 
$s_{\pm}$-wave state with the original SC transition temperature $T_{\rm c0} = 30$~K is completely suppressed
 when the residual resistivity reaches $\sim 5 z^{-1} \mu \Omega$cm, \cite{Onari_imp, Yamakawa_imp}
 where $z^{-1} = m^{*} / m$ is the mass-enhancement factor due to the self-energy. 
When $I^{\rm imp} = +1$~eV, the residual resistivity for $n^{\rm imp} = 0.01$ is about $20 \mu \Omega$cm in Fe-based superconductors. 
Therefore, the $s_{\pm}$-wave state is very fragile against impurity.

\section{Discussion}
\subsection{Violation of the Extinction Rule}
As shown in Sec.~\ref{sec-result}, the QPI peak around $\bm{q}_2$ is realized even in the $s_{++}$-wave state. 
The reason is that the numerator in Eq.~(\ref{eq-drodd-s}) has finite value
 under the experimental condition $E < | \Delta_{\rm h, e} |$.
Thus, the extinction rule in the octet model for cuprate superconductors
 $( | \Delta ( \bm{k}_i ) | = E < \Delta^{\rm max} ; i = 1 \sim 8)$,
 which tells that the QPI signal at $\bm{q} = \bm{k}_i - \bm{k}_j$ disappears
 if $\Delta ( \bm{k}_i ) = \Delta ( \bm{k}_j )$, 
 does not hold in Fe-based superconductors under the experimental condition $E < \Delta_{0}$. 
As shown in Fig.~\ref{fig3}, the QPI signal around $\bm{q}_2$ still exists even at $E = \Delta_{0}$
 in the $s_{++}$-wave state
 due to the finite quasiparticle damping $\gamma$. % is taken into account. 
For these reasons, we can not distinguish between the $s_{++}$-wave and $s_{\pm}$-wave states
 from the presence or absence of the QPI peak around $\bm{q}_2$.

\subsection{Comparison with Previous Studies}
In Ref.~\onlinecite{Sykora}, Sykora and Coleman investigated the QPI
 in the $s_{\pm}$-wave state by using a two-band model. 
They showed that the QPI peak around $\bm{q}_2$ emerges for $B = 0$
 due to the non-magnetic impurity scattering in the weak potential limit, 
 and its intensity is suppressed by the Zeeman effect under the magnetic field $B = \Delta_{0}$. 
It is consistent with the result of the present study for the weak potential case
 based on the five-orbital model. 
Also, to analyze the unitary scattering case,
 they phenomenologically treated the resonant scattering due to the multiple scattering process, 
 and proposed that the QPI signal around $\bm{q}_3 = ( \pi, \pi )$ is enhanced by $B$ due to the resonant scattering. 
However, we cannot obtain such behavior in the present study using $T$-matrix approximation
 for $I^{\rm imp} = +1$~eV. 

In Ref.~\onlinecite{Gao}, Gao {\it et al.} discussed the magnetic field dependence of the QPI due to the vortex,
 which is not considered in the present study. 
Interestingly, they showed that the strong and sharp QPI peak around $\bm{q}_3$ is caused 
 in both the $s_{++}$-wave and $s_{\pm}$-wave states
 by the Andreev scattering due to the vortices. 
Experimentally, however, the field-induced change is almost spatially uniform, 
 indicating that the impurity scattering is more important. \cite{Hanaguri}
In Ref.~\onlinecite{Gao}, the QPI peak around $\bm{q}_2$ was not obtained in the $s_{++}$-wave state
 maybe due to the very large difference in the band structure. 
 
\section{Summary}
In summary, we investigated the QPI in Fe-based superconductors
 in both the $s_{++}$-wave and $s_{\pm}$-wave states. 
In the octet model $\left( | \Delta ( \bm{k}_i ) | = E < \Delta^{\rm max}; i = 1 \sim 8 \right)$
 for cuprate superconductors with $d_{x^2 - y^2}$-wave SC state, 
 the QPI signal around $\bm{q} = \bm{k}_i - \bm{k}_j$ disappears
 when $\Delta ( \bm{k}_i )$ and $\Delta ( \bm{k}_j )$ have the same sign.
However, this extinction rule
 is not hold in Fe-based superconductors with fully-gapped $s$-wave SC state. 
The reason is that the resonance condition,
 in which the denominator of the integrand in Eq.~(\ref{eq-drodd-d}) becomes zero at some $\bm{k}$, 
 does not satisfied under the experimental condition $E < | \Delta_{\rm e, h} |$. 
We performed the detailed numerical study of the QPI signal on the basis of the five-orbital model and found that
 the experimentally observed QPI peak around $\bm{q}_2 = ( \pi, 0 )$
 can be explained in terms of both the $s_{++}$-wave and $s_{\pm}$-wave states. 
Furthermore, we discussed the magnetic field dependence of the QPI by considering the Zeeman effect,
 and found that the suppression of the peak intensity around $\bm{q}_2$ by the magnetic field can also be explained
 in terms of both the $s_{++}$-wave and $s_{\pm}$-wave states. 
Therefore, it is difficult to distinguish between the $s_{++}$-wave and $s_{\pm}$-wave states
 from the QPI experimental date for Fe-based superconductors.

\acknowledgments
We are grateful to T. Hanaguri for useful discussions. 
This study has been supported by Grants-in-Aid for Scientific Research from MEXT of Japan. 

\appendix

\section{QPI in Cuprate Superconductors}\label{sec-a1}
In the QPI measurement for the cuprate by Hanaguri {\it et al.}, \cite{Hanaguri_Cu}
 it was shown that the QPI signals due to the impurity scattering
 between $\bm{k}$ points with opposite sign gap functions
 are strongly suppressed by the magnetic field. 
Since the suppression in the ``matrix region" (far from vortex) is stronger
 than the one in the ``vortex region" (near the vortex core),
 the Zeeman effect would be important. 
In this appendix, we investigate the magnetic field dependence of the QPI
 in cuprate superconductors with nodal $d_{x^2 - y^2}$-wave SC state,
 $\Delta_{\bm{k}} = \Delta_0 ( \cos{k_x} - \cos{k_y} ) / 2$, 
 using the $T$-matrix approximation in the case of weak impurity potential $I^{\rm imp} = 0.1$~eV, 
 and show that the experimentally observed suppression can be explained by the Zeeman splitting scenario. 

%%%%%%%%%%%%%%%%%%%%%%%%%%%%%%%%%%%%%%%%
\begin{figure}[tbp]
	\includegraphics[width=1.0\linewidth]{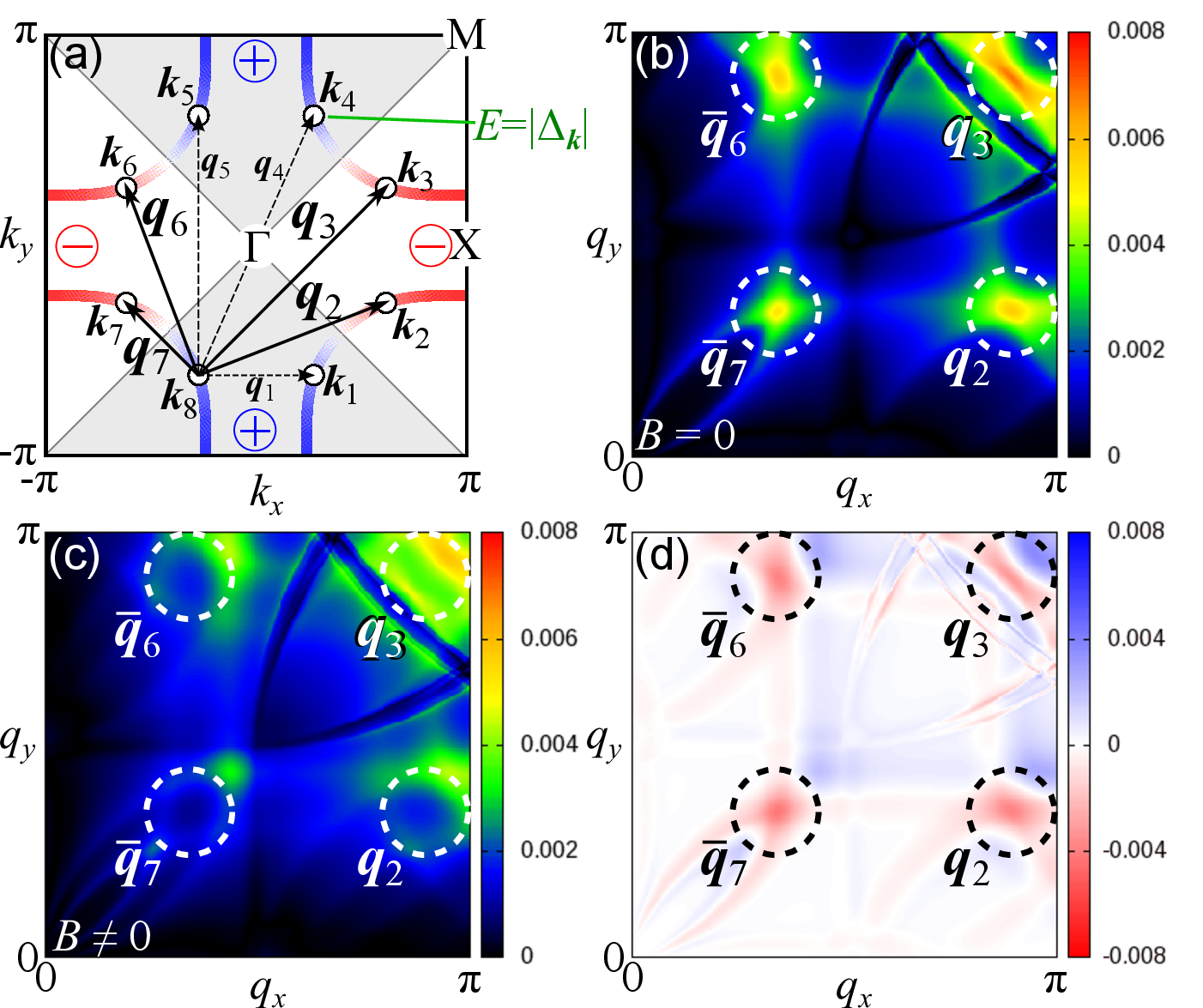}
	\caption{(Color online)
		(a) FS of cuprate and the $d_{x^2-y^2}$-wave SC gap. 
		The wave vector $\bm{k}_i$ ($i = 1 \sim 8$) satisfies the relation $ E = | \Delta_{\bm{k}_i} |$,
		 and $\bm{q}_i$ is scattering vector. 
		$\bm{q}_{2, 3, 6, 7}$ ($\bm{q}_{1, 4, 5}$) connect the $\bm{k}$-points on FS
		 with opposite (same) sign gap functions. 
		(b) Intensity map of $| Z ( \bm{q}, E ) |$ at $E = \Delta_0 / 2$ and $B$ = 0. 
		The scattering vector $\bar{\bm{q}}_i$ is equivalent to $\bm{q}_i$. 
		(c) $| Z ( \bm{q}, E ) |$ for $B = \Delta_0 / 8$. 
		(d) Field-induced change given by $| Z ( \bm{q}, E ) |_{B = \Delta_0 / 8} - | Z ( \bm{q}, E ) |_{B = 0}$.
	}
	\label{figa1}
\end{figure}
%%%%%%%%%%%%%%%%%%%%%%%%%%%%%%%%%%%%%%%%

Figure~\ref{figa1}(a) shows the FS and the gap function in cuprate superconductors. 
The eight wave vectors $\bm{k}_i$ ($i = 1 \sim 8$) on the FS
 satisfy the relation $ E = | \Delta_{\bm{k}_i} | < \Delta_{0}$. 
The scattering vectors $\bm{q}_{1, 4, 5}$ ($\bm{q}_{2, 3, 6, 7}$) connect the two $\bm{k}$-points
 with same (opposite) sign gap functions. 
Experimentally, the QPI signals are obtained at $\bm{q}_{2, 3, 6, 7}$ for zero field,
 and they are suppressed by applying a magnetic field. \cite{Hanaguri_Cu}
Figure~\ref{figa1}(b) shows the numerical results
 of the QPI intensity map $| Z ( \bm{q}, E ) |_{B = 0}$ without magnetic field. 
We use the parameters given in Ref.~\onlinecite{Maltseva}. 
The strong QPI peaks appear at $\bm{q}_{2, 3, 6, 7}$. 
Figures~\ref{figa1}(c) and (d) show the QPI with magnetic field, $| Z ( \bm{q}, E ) |_{ B = \Delta_{0} / 8}$,
 and field-induced change
 given by $| Z ( \bm{q}, E ) |_{B = \Delta_{0} / 8} - | Z ( \bm{q}, E ) |_{ B = 0}$, respectively. 
In this case, the QPI signal shows remarkable field dependence
 and its peaks at $\bm{q}_{2, 3, 6, 7}$ are strongly suppressed by the Zeeman effect. 
This result is consistent with the experimental results for cuprate superconductors. \cite{Hanaguri_Cu}

\section{QPI due to Simplified Impurity Potential}
In the above discussion, we have investigated the QPI
 due to the orbital diagonal impurity potential in Eq.~(\ref{eq:iimp}). 
In this case, the impurity potential has complex $\bm{k}$-dependence in the band basis. 
In this appendix, we consider the QPI due to a simple
 constant impurity potential in the band basis,
\begin{eqnarray}\label{eq:iband}
	I^{\rm band}_{b, b'}
  =	\left\{
	\begin{array}{cc}
		I  & ( b = b' ) \\
		I' & ( b \ne b' )
	\end{array}
	\right.
,
\end{eqnarray}
 where $b = b'$ and $b \ne b'$ terms correspond to intraband and interband scattering, respectively. 
Hereafter, we study the QPI in the weak potential case with $I = I' = 0.1$~eV. 

%%%%%%%%%%%%%%%%%%%%%%%%%%%%%%%%%%%%%%%%
\begin{figure}[tbp]
 	\includegraphics[width=1.0\linewidth]{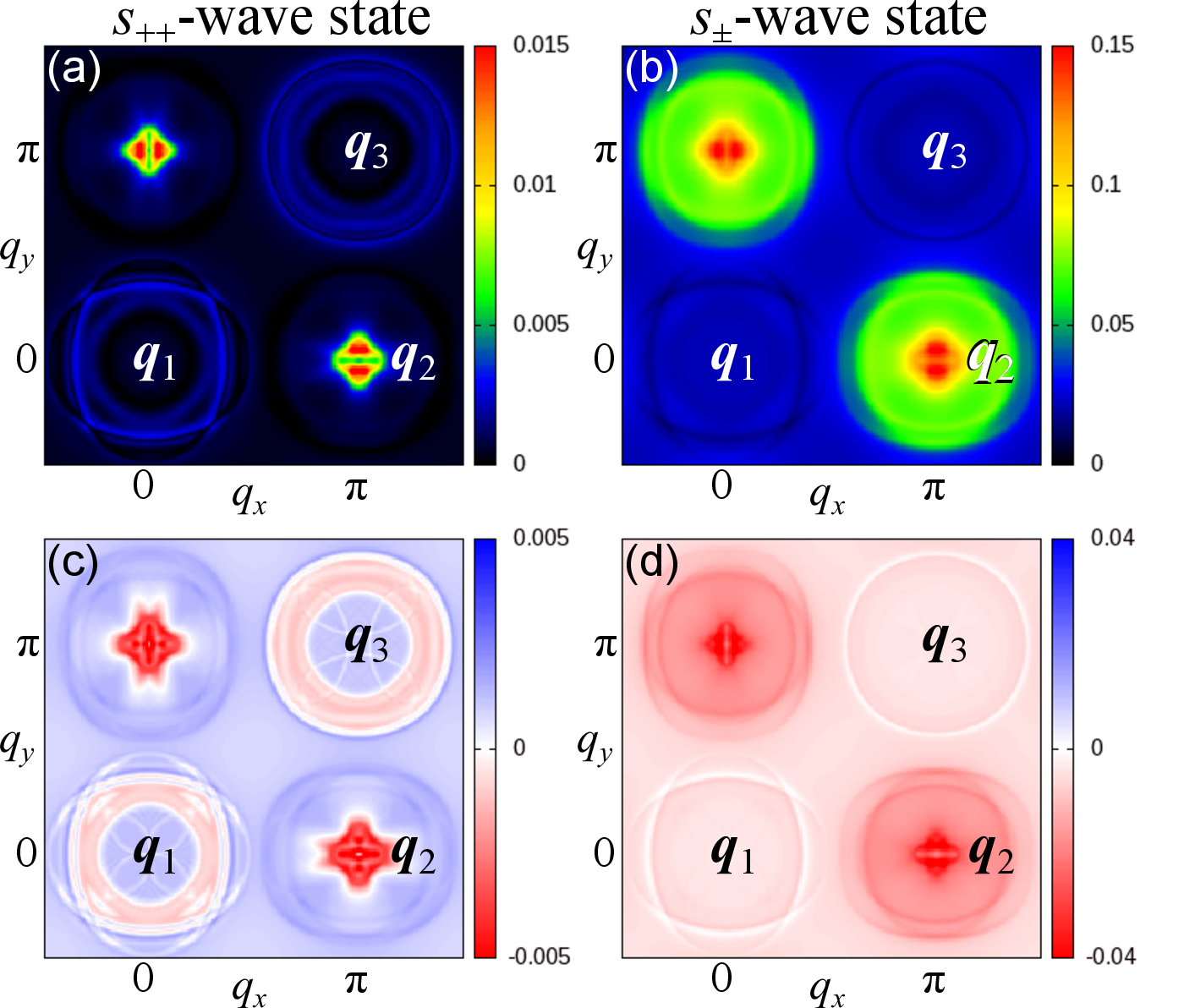}
	\caption{(Color online)
		(a)(b) Intensity map of the QPI $| Z ( \bm{q}, E ) |_{B = 0}$ 
		 due to the band represented weak impurity potential $I^{\rm band}_{b,b'} = 0.1$~eV 
		 in the (a) $s_{++}$-wave and (b) $s_{\pm}$-wave states. 
		(c)(d) The magnetic field-induced change of the QPI signal
		 given by $| Z ( \bm{q}, E ) |_{B = \Delta_{0} / 2} - | Z ( \bm{q}, E ) |_{0}$
		 in the (c) $s_{++}$-wave and (d) $s_{\pm}$-wave states. 
		We set $E = 0.7 \Delta_{0}$, $\Delta_{\rm h} = 2 \Delta_{0}$ and $\Delta_{\rm e} = \pm \Delta_{0}$. 
	}
	\label{figa2}
\end{figure}
%%%%%%%%%%%%%%%%%%%%%%%%%%%%%%%%%%%%%%%%
Figures~\ref{figa2}(a) and (b) show the QPI intensity map 
in the $s_{++}$-wave and $s_{\pm}$-wave states, respectively. 
We set $E = 0.7 \Delta_0$, $ | \Delta_{\rm h} | = 2 \Delta_0 $, and $ | \Delta_{\rm e} | = \Delta_0 $. 
The QPI peak around $\bm{q}_2$ appears in both the $s_{++}$-wave and $s_{\pm}$-wave states. 
Figures~\ref{figa2}(c) and (d) show the field-induced change
 $| Z ( \bm{q}, E ) |_{B = \Delta_{0} / 2} - | Z ( \bm{q}, E ) |_{B = 0}$
 in the $s_{++}$-wave and $s_{\pm}$-wave states, respectively. 
The obtained results are qualitatively consistent
 with the orbital diagonal potential case
 shown in Figs.~\ref{fig2}(c), \ref{fig2}(d), \ref{fig3}(c), and \ref{fig3}(d) in the main text.
Therefore, in the weak potential case, 
the obtained QPI signal is insensitive
 to the nature of impurity potential. 

However, the impurity potential in Eq.~(\ref{eq:iband}) gives an erroneous result
 in the unitary regime $I \to \infty$,
 that is, the $T$-matrix $T^{\rm band}_{b, b'}$ becomes band diagonal except for $I' / I = 1$. 
Due to this model artifact, 
 the QPI peak around $\bm{q}_2$ disappears in the unitary limit. 
For the same reason, $T_{\rm c}$ in the $s_{\pm}$-wave state is almost unchanged
 by impurities in the unitary regime $I \rho^{0} (0) \ge 1$. \cite{Senga, Wang}
However, such erroneous model artifact is revised by using a realistic potential
 in Eq.~(\ref{eq:iimp}). \cite{Onari_imp, Yamakawa_imp}
That is, the QPI peak around $\bm{q}_2$ appears and $T_{\rm c}$ in the $s_{\pm}$-wave state
 is fragile against impurity even in the unitary regime.

\section{Another Two-Gap Case with $| \Delta_{\rm h} | = 1.5 | \Delta_{e} |$}
In the main text, 
 we discussed the field-induced suppression of the QPI peak intensity around $\bm{q}_2$ 
 in the isotropic two-gap case with $| \Delta_{\rm h} | = 2 | \Delta_{\rm e} |$. 
Here, we show another two-gap case with $| \Delta_{\rm h} | = 1.5 |\Delta_{\rm e} |$. 
The obtained results are qualitatively
 the same as the results for $| \Delta_{\rm h} | = 2 | \Delta_{\rm e} |$ in the main text. 

\begin{figure}[htbp]
	\includegraphics[width=0.45\linewidth]{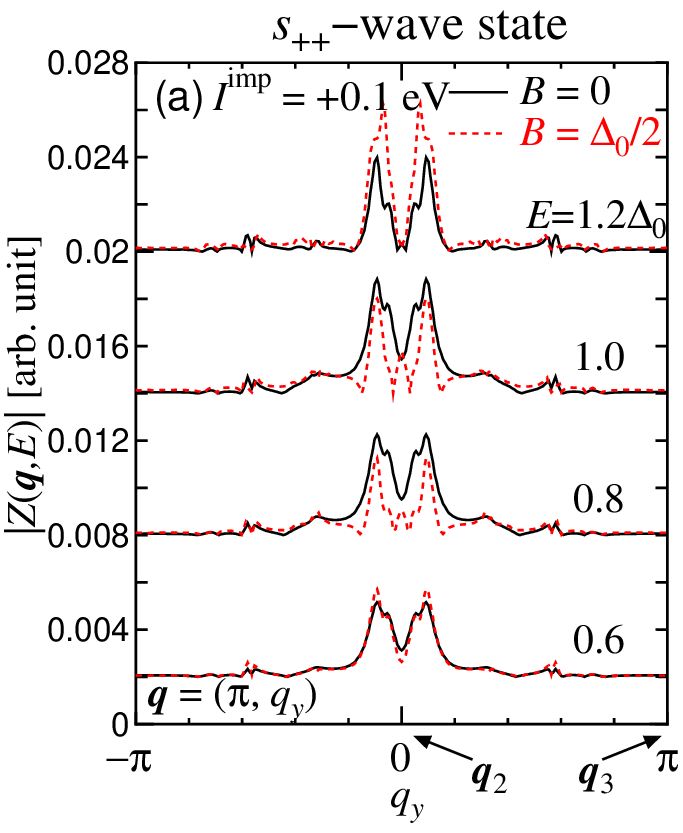}
	\includegraphics[width=0.45\linewidth]{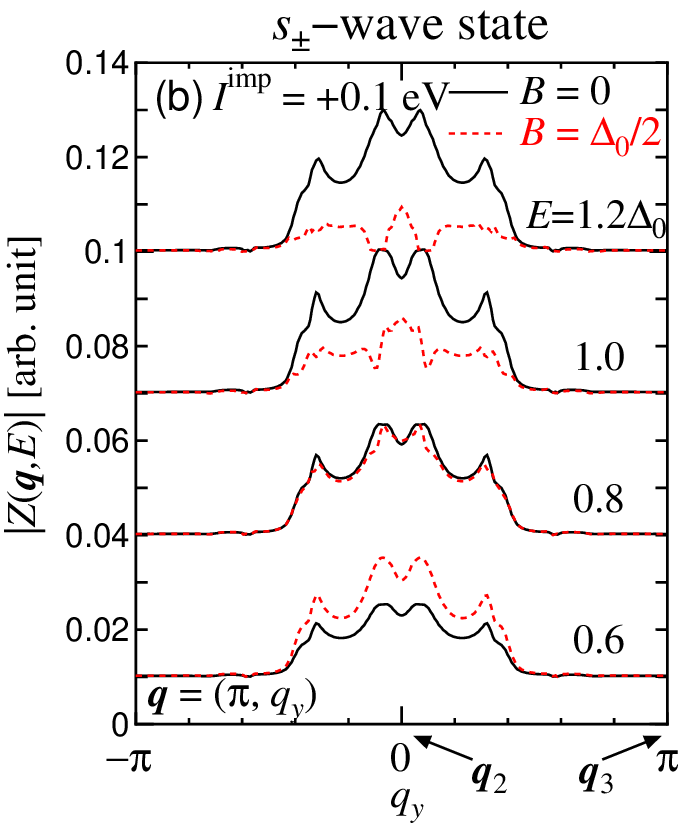}
	\includegraphics[width=0.45\linewidth]{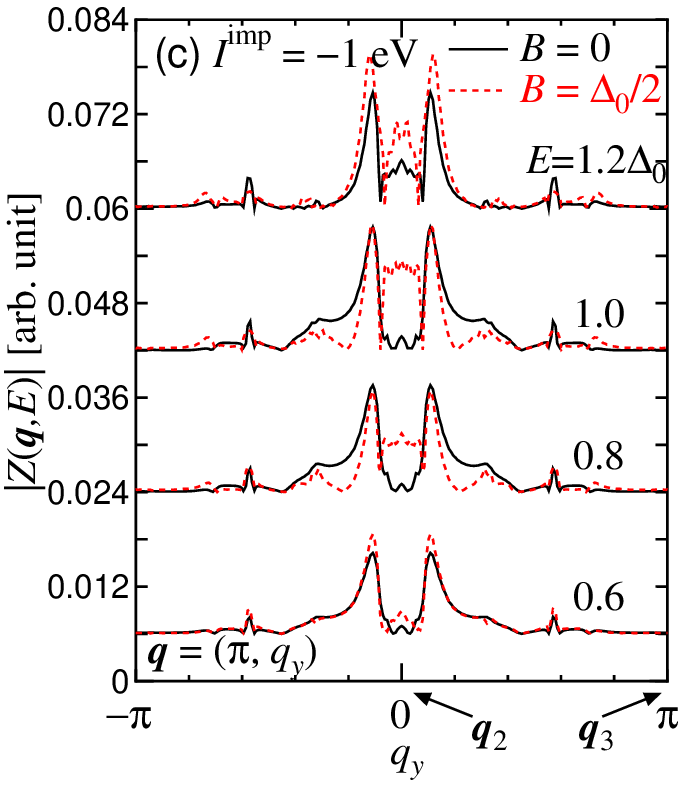}
	\includegraphics[width=0.45\linewidth]{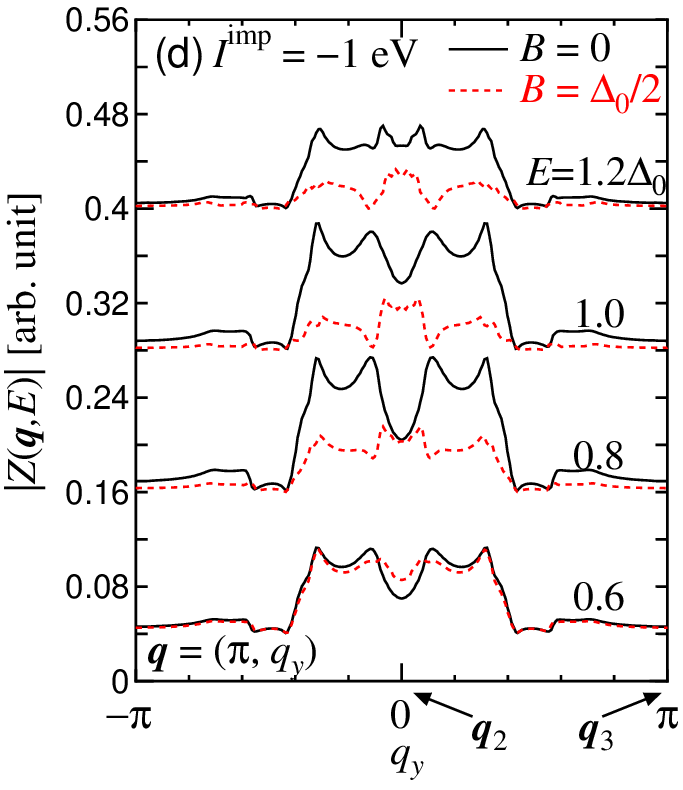}
	\includegraphics[width=0.45\linewidth]{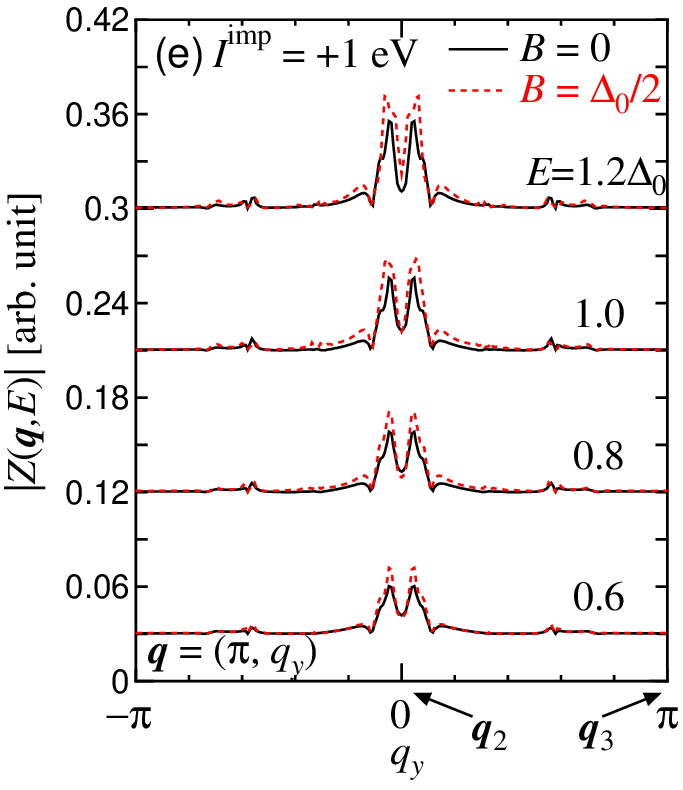}
	\includegraphics[width=0.45\linewidth]{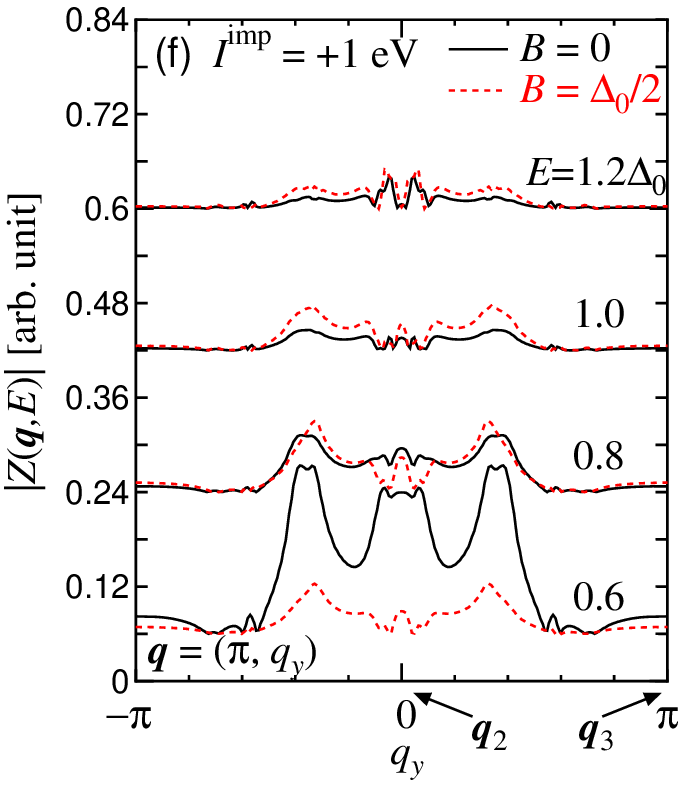}
	\caption{(Color online)
		$| Z ( \bm{q}, E ) |$ along $\bm{q} = (\pi, q_y )$
		 in the isotropic two-gap case with $\Delta_{\rm h} = 1.5 \Delta_{0}$
		 and $\Delta_{\rm e} = \pm \Delta_{0}$. 
		The solid and dotted lines represent $B = 0$ and $B = \Delta_{0} / 2$, respectively. 
		The left and right panels show the results
		 in the $s_{++}$-wave and $s_{\pm}$-wave states, respectively. 
		(a),(b) $I^{\rm imp} = 0.1$ eV.
		(c),(d) $I^{\rm imp} = - 1$ eV.
		(e),(f) $I^{\rm imp} = + 1$ eV.
		The curves in all the figures are vertically shifted to make them visible. 
	}
	\label{figs3}
\end{figure}

Figure~\ref{figs3} shows the $| Z ( \bm{q}, E ) |$ from $\bm{q} = ( \pi, -\pi )$
 to $( \pi, \pi )$ for $B = 0$ (solid lines) and $B = \Delta_{0} / 2$ (dotted lines). 
In the (a) $s_{++}$-wave and (b) $s_{\pm}$-wave states,  
 the QPI intensity for $I^{\rm imp} = 0.1$~eV around $\bm{q}_2$ is suppressed by $B$ for $E \sim \Delta_{0}$. 

Figure~\ref{figs3}(c) shows the $| Z ( \bm{q}, E ) |$ in the $s_{++}$-wave state for $I^{\rm imp} = -1$~eV. 
In this case, the QPI intensity at just $\bm{q}_2$ is strongly enhanced by $B$ at $E \sim \Delta_{0}$,
 whereas the integrated intensity around $\bm{q}_2$ is suppressed. 
Such field-induced enhancement at just $\bm{q}_2$ for $E = \Delta_0$ is not universal
 since the $\bm{q}_2$ peak is suppressed by $B$ for $| \Delta_{\rm h} | = 2 | \Delta_{\rm e} |$
 as shown in Fig. \ref{fig5}(a) in the main text. 
However, the obtained field-induced enhancement at just $\bm{q}_2$ may be consistent with the experimental result. 
Experimentally, the QPI signal for $E = 1.0$~meV is suppressed by $B$ around $\bm{q}_2$,
 but a slight enhancement is observed at just $\bm{q}_2$ as shown in Fig.~1(A) in Ref. \onlinecite{Hanaguri_arxiv}. 

Figure~\ref{figs3} (d) shows the $| Z ( \bm{q}, E ) |$ in the $s_{\pm}$-wave state for $I^{\rm imp} = -1$~eV. 
Also, Figs.~\ref{figs3}(e) and (f) show the ones for $I^{\rm imp} = +1$~eV.
In all cases (d)-(f) in Fig.~\ref{figs3},
 the obtained results are almost same as the
 cases (b)-(d) in Fig.~\ref{fig5} in the main text for $| \Delta_{\rm h} | = 2 |\Delta_{\rm e} |$. 

Therefore, the obtained results for $| \Delta_{\rm h} | = 1.5 |\Delta_{\rm e} |$
 are qualitatively same as the ones for $| \Delta_{\rm h} | = 2 | \Delta_{\rm e} |$ in the main text. 
The field-induced enhancement at just $\bm{q}_2$ 
 for $I^{\rm imp} = - 1$~eV in Fig.~\ref{figs3}(c)
 may be consistent with experimental result,
 although it is sensitive to model parameters.

%%%%%%%%%%%%%%%%%%%%%%%%%%%%%%%%%%%%%%%%
%%%%%%%%%%%%%%%%%%%%%%%%%%%%%%%%%%%%%%%%

\end{document}